\documentclass[11pt,a4paper]{article}
\usepackage{a4wide, axodraw}
\usepackage{cite}
\usepackage{amsmath,amsthm,amssymb,amsfonts}
\usepackage{array}
\makeatletter
\renewcommand\section{\@startsection {section}{1}{\z@}%
                                   {-3.5ex \@plus -1ex \@minus -.2ex}
                                   {2.3ex \@plus.2ex}%
                                   {\normalfont\large\bfseries}}
\renewcommand\subsection{\@startsection{subsection}{2}{\z@}%
                                     {-3.25ex\@plus -1ex \@minus
                                     -.2ex}%
                                     {1.5ex \@plus .2ex}%
                                     {\normalfont\bfseries}}

\begin{document}

\newcommand{\eg}{\emph{e.g.}\;}
\newcommand{\ie}{\emph{i.e.}}
\newcommand{\etal}{\emph{et al.}}
\newcommand{\nn}{\nonumber}
\newcommand{\Dslash}{\not{\hbox{\kern-4pt $D$}}}
\newcommand{\Supertwistor}{\Cset \mathrm{P}^{3|4}}
\newcommand{\Twistorspace}{\Cset \mathrm{P}^{3}}
\newcommand{\MHV}{\mathbb{C} \mathrm{P}^{1}}
\newcommand{\AdS}{\mathrm{AdS}}
\newcommand{\half}{\frac{1}{2}}
\newcommand{\diff}{\mathrm{d}}
\newcommand{\ra}{\rightarrow}
\newcommand{\Zset}{{\mathbb Z}}
\newcommand{\Cset}{{\,\,{{{^{_{\pmb{\mid}}}}\kern-.47em{\mathrm C}}}}}
\newcommand{\gra}{\alpha}
\newcommand{\grl}{\lambda}
\newcommand{\gre}{\epsilon}
\newcommand{\zb}{{\bar{z}}}
\newcommand{\mn}{{\mu\nu}}
\newcommand{\Acal}{{\mathcal A}}
\newcommand{\Rcal}{{\mathcal R}}
\newcommand{\Dcal}{{\mathcal D}}
\newcommand{\Mcal}{{\mathcal M}}
\newcommand{\Ncal}{{\mathcal N}}
\newcommand{\Kcal}{\mathcal{K}}
\newcommand{\Lcal}{{\mathcal L}}
\newcommand{\Scal}{{\mathcal S}}
\newcommand{\Wcal}{{\mathcal W}}
\newcommand{\Bcal}{\mathcal{B}}
\newcommand{\Ccal}{\mathcal{C}}
\newcommand{\Jcal}{\mathcal{J}}
\newcommand{\Vcal}{\mathcal{V}}
\newcommand{\Ocal}{\mathcal{O}}
\newcommand{\Qcal}{\mathcal{Q}}
\newcommand{\Zcal}{\mathcal{Z}}
\newcommand{\Zb}{\overline{Z}}
\newcommand{\Urm}{{\mathrm U}}
\newcommand{\Srm}{{\mathrm S}}
\newcommand{\SO}{\mathrm{SO}}
\newcommand{\Sp}{\mathrm{Sp}}
\newcommand{\SU}{\mathrm{SU}}
\newcommand{\SL}{\mathrm{SL}}
\newcommand{\U}{\mathrm{U}}
\newcommand{\PSU}{\mathrm{PSU}}
\newcommand{\be}{\begin{equation}}
\newcommand{\ee}{\end{equation}}
\newcommand{\bea}{\begin{eqnarray}}
\newcommand{\eea}{\end{eqnarray}}
\newcommand{\tQ}{\tilde{Q}}
\newcommand{\trho}{\tilde{\rho}}
\newcommand{\tphi}{\tilde{\phi}}
\newcommand{\lt}{\tilde{\lambda}}
\newcommand{\dagphi}{{\phi^\dagger}}
\newcommand{\dagq}{{q^\dagger}}
\newcommand{\dagz}{{z^\dagger}}
\newcommand{\bzeta}{{\bar{\zeta}}}
\newcommand{\blambda}{{\bar{\lambda}}}
\newcommand{\bchi}{{\bar{\chi}}}
\newcommand{\tmu}{\tilde{\mu}}
\newcommand{\mut}{\tilde{\mu}}
\newcommand{\ad}{\dot a}
\newcommand{\dbar}{\bar{\partial}}
\newcommand{\Tr}{\mathrm{Tr}}
\newcommand{\Comment}[1]{{}}
\newcommand{\doublet}[2]{\left(\begin{array}{c}#1\\#2\end{array}\right)}
\newcommand{\twobytwo}[4]{\left(\begin{array}{cc} #1&#2\\#3&#4\end{array}\right)}
\newcommand{\note}[2]{{\footnotesize [{\sc #1}}---{\footnotesize   #2]}}
\newcommand{\Dc}{\mathrm{D}_c}
\newcommand{\Df}{\mathrm{D}_f}
\newcommand{\Zbar}{\bar{Z}}
\newcommand{\p}{\partial}

\makeatletter
\@addtoreset{equation}{section}
\makeatother
\renewcommand{\theequation}{\thesection.\arabic{equation}}

\rightline{QMUL-PH-07-13} \rightline{CERN-PH-TH/2007-117}
\rightline{TIFR/TH/07-13} \vspace{2truecm}

\vspace{15pt}


\centerline{\LARGE \bf Twistor Strings with Flavour} \vspace{1truecm}
\thispagestyle{empty} \centerline{
    {\large \bf James Bedford${}^{a,b,}$}\footnote{E-mail address:
                                  {\tt james.bedford@cern.ch}},
    {\large \bf Constantinos Papageorgakis${}^{c,}$}\footnote{E-mail address:
                                  {\tt costis@theory.tifr.res.in}}
    {\bf and}
    {\large \bf Konstantinos Zoubos${}^{a,}$}\footnote{E-mail address:
                                  {\tt k.zoubos@qmul.ac.uk}}
                                                       }

\vspace{.4cm}
\centerline{{\it ${}^a$ Centre for Research in String Theory, Department of Physics}}
\centerline{{ \it Queen Mary, University of London}} \centerline{{\it Mile End Road, London E1 4NS, UK}}

\vspace{.4cm}
\centerline{{\it ${}^b$ Department of Physics, CERN - Theory
    Division}}
\centerline{{\it 1211 Geneva 23, Switzerland}}

\vspace{.4cm}
\centerline{{\it ${}^c$ Department of Theoretical Physics, Tata
    Institute of Fundamental Research}} \centerline{{\it Homi Bhabha
    Road, Mumbai 400 005, India}}

\vspace{1.5truecm}

\thispagestyle{empty}

\centerline{\bf ABSTRACT}

\vspace{.5truecm}

\noindent We explore the tree--level description of a class of  $\Ncal = 2$ UV-finite
SYM theories with fundamental flavour within a topological B--model twistor string
framework. In particular, we identify the twistor dual of the $\Sp(N)$ gauge theory
with one antisymmetric and four fundamental hypermultiplets, as well as that of the
$\SU(N)$ theory with $2N$ hypermultiplets. This is achieved by suitably
orientifolding/orbifolding the
original $\Ncal = 4$  setup of Witten and adding a certain number of
new topological `flavour'--branes at the orientifold/orbifold fixed
planes to provide the fundamental matter. We further comment on the
appearance of these objects in the B--model on $\Supertwistor$.
An interesting aspect of our construction is that, unlike the IIB description of these
theories in terms of D3 and D7--branes, on the twistor side part of the global flavour
symmetry is realised geometrically. We provide evidence for this correspondence by
calculating and matching amplitudes on both sides.

\vspace{.5cm}

\setcounter{page}{0}

\newpage

\tableofcontents

\setcounter{footnote}{0}


\section{Introduction}

Four--dimensional conformal field theories are relatively rare, and their
existence depends crucially on the presence of a large amount of symmetry. The
most celebrated example is $\Ncal = 4$ supersymmetric Yang--Mills (SYM) theory,
which, especially via its strong--weak duality with IIB string theory on
$\AdS_5\times \Srm^5$ \cite{Maldacena97}, has provided a very useful
testbed for understanding
the physics of strongly-coupled gauge theory. In this duality, the exact quantum conformal
invariance of the theory is reflected in the $\AdS_5$ factor of the string background,
which encodes the unbroken four--dimensional conformal group $\SO(2,4)$ of the gauge theory.

 A very different duality involving $\Ncal = 4$ SYM was proposed by Witten in 2003
\cite{Witten0312}. The idea stems from the fact that certain scattering amplitudes in Yang--Mills theory,
when expressed in appropriate (spinor helicity) variables, turn out
to take an unexpectedly simple form. This indicates that there might exist some reformulation
in which this simplicity is evident, and in this context Witten proposed that it is useful to
consider the open-string topological
B--model on supertwistor space $\Supertwistor$. The isometries of $\Supertwistor$ capture
the superconformal group $\PSU(2,2|4)$ of the gauge theory, and the spectrum of the string
theory can be mapped to the field content of $\Ncal = 4$ SYM via the Penrose transform \cite{Penrose67}.

In this framework, gluon scattering amplitudes can be calculated by noting that they are supported
on certain simple algebraic curves in twistor space, the degree of which is linked to the number
of external negative helicity gluons. For instance, Maximally Helicity Violating
(MHV) amplitudes,  which have two negative and any number of positive helicity gluons,
are supported on degree one curves in $\Supertwistor$. In \cite{Witten0312} it was
proposed that these curves are wrapped by D1--instantons in the B--model, and,
adapting a method originally due to Nair \cite{Nair88}, it was shown that appropriately
integrating over the moduli space of these D1--instantons leads to the correct
expressions for tree-level amplitudes in $\Ncal = 4$ SYM.

Beyond tree level, however, the situation is very different. Apart from difficulties in understanding the appropriate measure
for higher--genus curves in supertwistor space, at one loop it seems that one cannot
avoid unwanted contributions from the closed B--model sector which
would correspond to conformal supergravity states in spacetime \cite{BerkovitsWitten04}. As the action for
conformal supergravity is the square of the Weyl tensor whose kinetic term is fourth order in
derivatives, it is generally believed to be non-unitary and thus a highly undesirable feature. Nonetheless,
loop amplitudes in such a theory have been investigated \cite{DolanGoddard0703} using an alternative twistor string theory
due to Berkovits \cite{Berkovits0402} and it is hoped that one might still be able to learn something
about loop amplitudes in Yang-Mills this way.

Despite the above shortcoming, the application of twistor-inspired techniques to gauge theory has
resulted in great progress in the understanding of perturbative field theory. At tree-level,
the realisation that amplitudes localising on degree $d$ curves can be equivalently
calculated by integrating over the moduli space of $d$ disconnected degree 1 curves
\cite{Roibanetal0402,RoibanVolovich0402,Roibanetal0403,Cachazoetal0403},
underlies the so-called MHV (or CSW) rules proposed by Cachazo, Svr\v{c}ek and Witten
\cite{Cachazoetal0403}. The CSW rules elevate tree-level MHV amplitudes to effective
vertices, which
are then glued together using simple scalar propagators to form tree amplitudes with
successively greater numbers of negative helicity particles. Of particular interest is
the fact that these techniques are applicable to a far larger class of theories than
$\Ncal = 4$ SYM, and include gauge theories with reduced or no supersymmetry and
Einstein (super-)gravity - see
\cite{Cachazoetal0403,GeorgiouKhoze0404,Georgiouetal0407,DixonGloveretal0411,Bohretal0509} and references therein.

Even more remarkable is the fact that, despite the apparent failure of the twistor string
duality at loop--level, the MHV rules \emph{can} be straightforwardly applied at one loop
in $\Ncal = 4$ SYM \cite{Brandhuberetal0407}, $\Ncal = 2,1$ SYM \cite{Bedfordetal04,QuigleyRozali04},
pure YM \cite{Bedfordetal0412}, a certain effective Higgs--YM action \cite{Badgeretal07} and
$\Ncal\!=\!8$ supergravity \cite{NastiTravaglini07}.
These results would seem to indicate that it is possible to overcome the current difficulties
at one-loop and eventually extend Witten's prescription to the quantum level
not only for $\Ncal = 4$ SYM, but also for the other theories above. It is
possible that such a dual string theory would have to be an appropriate
(non--topological?) extension of the B--model,
perhaps combined with a modification of the bosonic part of the target space geometry away from
$\Twistorspace$ to reflect the fact that conformal invariance is typically lost at
the quantum level. Finding such a quantum completion of the twistor string framework
would certainly deepen our understanding of perturbative gauge theory.

As an intermediate step towards this goal, it is important to map out the range of
four--dimensional theories that can potentially admit a twistor string description.
If, to restrict the question somewhat, we insist that the full quantum theory have
a perturbative string dual containing twistor space as part of the target manifold,
we should clearly look among the known quantum conformally invariant theories, and,
if we require that the conformal symmetry holds order--by--order in the
coupling, we should focus in particular on the subset of the above which are finite.
The hope is that, by explicitly constructing the twistor string duals of a wide
range of such theories, which are expected to retain $\Twistorspace$ as part of
the geometry at loop level, and by understanding why this construction might not
work for other theories which look similar classically but which lack conformal
invariance at the quantum level, one may learn something about the properties
of the elusive quantum twistor string. In the process, one might also hope to
gain further insight into the B--model twistor string description (or any of the
several alternatives \cite{Berkovits0402,Siegel0404}) even at tree--level.

Following this programme, it was shown in \cite{KulaxiziZoubos04} (see also \cite{GaoWu06}) that
the $\Ncal = 1$ exactly marginal deformations of $\Ncal = 4$ SYM can be incorporated
into the B--model description by turning on a particular closed string mode, which
(via a certain open/closed correlation function) effectively introduces
non--anticommutativity between some of the fermionic coordinates of $\Supertwistor$.
Another class of known finite 4d gauge theories are the quiver theories that arise
as $\Ncal = 1$ and $\Ncal = 2$ orbifolds of $\Ncal = 4$ SYM and in \cite{ParkRey04,Giombietal04}
it was shown that these theories also admit a very natural twistor string
description.\footnote{Twistor string duals have also been constructed for
truncations of self--dual $\Ncal=4$ SYM \cite{PopovWolf04}, lower dimensions
\cite{Chiouetal0502, Popovetal05,Saemann05,LechtenfeldSaemann05,Popov:2007hb}, chiral mass terms
\cite{Chiouetal0512}
as well as for a number of gravity theories including $\Ncal = 1,2$ conformal
supergravity \cite{Ahn0409,Ahn0412} and Einstein supergravity \cite{AbouZeidHull0606}.}

In the present work we extend this investigation to other types of 4d gauge
theories by  including matter transforming in the fundamental representation.
These are the $\Ncal = 2$ SYM theories with gauge groups $\Sp(N)$ and $\SU(N)$,
which are UV--finite when the number of flavours is $N_f = 4$ and $N_f\! =\! 2N$ respectively, and where the
$\Sp(N)$ theory also contains a hypermultiplet in the antisymmetric representation.
For brevity we will refer to these simply as  the
$N_f = 4$ and $N_f\! =\! 2N$ theories.

In direct analogy with the stringy description of the $N_f =4$ gauge theory,
in order to obtain a symplectic gauge group it will be necessary
to perform an orientifold of the B--model on $\Supertwistor$.
Similarly,  for the $N_f\! =\! 2N$ theory we will perform
an orbifold projection. Given the similarities of these techniques
with previous orbifold constructions of \cite{ParkRey04,Giombietal04},
the above steps are relatively
straightforward. The main novelty, compared to the previous twistor string literature, is
the presence of the fundamental flavours. We propose a natural mechanism to
incorporate this sector of the theory, leading to an additional term in the B--model
action, and show that the tree--level twistor string amplitudes precisely match
those calculated on the gauge theory side.

A parallel promising development in the twistor string programme has been the
introduction of effective actions on twistor space \cite{Boelsetal06,Boelsetal07,Boels07},
which extend Witten's holomorphic Chern--Simons (hCS) action and, after appropriate
gauge fixing, reproduce the 4d MHV--rules prescription for Yang--Mills
theory.\footnote{Recently, some aspects of this formalism were extended to
self--dual $\Ncal=8$ supergravity \cite{MasonWolf07}.} By construction, this approach
does not suffer from the conformal supergravity problem. It is not yet known whether such
actions can be derived from a more fundamental (B--model or alternative) string
description (in particular, they do not seem to arise from simple summation over the
effects of D--instantons). Such an effective action for
either of the $\Ncal=2$ theories that we will consider in this work, constructed by inserting
the relevant matter multiplets (as described in \cite{Boelsetal06}),
and choosing the gauge group appropriate for each case, would provide an
alternative way to reproduce the MHV amplitudes we will calculate.
However, we do not follow that path since via such an approach we would not
expect to gain insight into the novel features
that arise when introducing fundamental flavours from a \emph{topological string}
point of view. Nevertheless, as we will point out, some aspects of our construction
will turn out to be similar to those in \cite{Boelsetal06}.

The rest of this paper is organised as follows:
In Section 2 we discuss some preliminary details related to formulating
the spacetime action for the $N_f = 4$ theory.
We then review Witten's construction of the twistor string for $\Ncal = 4$ SYM
and proceed to give the equivalent description for the theory under present study in Section 3.
In Section 4 we elaborate on the comparison between amplitudes calculated from the
spacetime and twistor points of view and demonstrate the agreement between the
two pictures with a number of specific examples.
Section 5  extends the above to the case with $N_f\!=\!2 N$.
We describe the construction of  the spacetime action, obtain the dual twistor string
description and finally match the two by comparing amplitude ratios.
We conclude in Section 6 with a discussion of our results and directions for future research.

\section{Preliminaries for the $N_f\! =\!4$ theory} \label{Nf4sec}

 The aim of this section is to collect known facts on the $N_f=4$ theory and
it symmetries, before moving on to considering its spacetime action. It is
easy to check that the matter content of this $\Ncal=2$, $\Sp(N)$ theory
(one hypermultiplet in the antisymmetric representation of $\Sp(N)$ and four
hypermultiplets in the fundamental) is such that the one--loop $\beta$--function
vanishes \cite{Howeetal83,ParkesWest84}.
Since $\Ncal=2$ supersymmetry implies one--loop exactness of the $\beta$--function
\cite{GrisaruSiegel82},
perturbative finiteness is guaranteed. In the rank one case, where the
gauge group reduces to $\SU(2)$ and there is no antisymmetric hypermultiplet,
this theory was considered by Seiberg
and Witten \cite{SeibergWitten9407,SeibergWitten9408}, who found (for arbitrary
hypermultiplet masses) the curve describing
its low energy dynamics. In the massless case, these results can be used to argue
that the gauge coupling does not run even at the nonperturbative
level.\footnote{Certain discrepancies in matching the results of \cite{SeibergWitten9408}
to explicit instanton calculations were resolved
in \cite{Doreyetal9611}.} One of the intriguing outcomes of \cite{SeibergWitten9408}
was the conjecture that the $N_f=4$ theory enjoys an analogue of the
Montonen--Olive (electric-magnetic) duality of $\Ncal=4$ SYM, in which $\SL(2,\Zset)$
mixes in a nontrivial way with $\SO(8)$ triality to produce a duality--invariant
spectrum.

\subsection{Review of the IIB/F-theory embedding}\label{ftheorysec}

The $N_f = 4$ theory has a very useful realisation in terms of a physical string theory
description, which first arose in Sen's explorations of F-theory
\cite{Vafa9602} on $K3$ \cite{Sen9605}. In particular, Sen considered a special
elliptically fibred $K3$, the orbifold  $T^4/\mathbb Z_2$, realised as a $T^2$
fibration over the base $T^2/\mathbb Z_2$. Requiring that the axion--dilaton modulus
have no dependence on the internal torus, this configuration reduces to an
orientifold \cite{GimonPolchinski96} of type IIB\footnote{For related reviews on
  brane dynamics in the presence of orientifolds see \eg \cite{Dabholkar98,GiveonKutasov98}.} on $T^2$, and thus produces four
orientifold fixed planes, each carrying $-4$ units of D7--brane charge. Constancy of
the axion--dilaton requires that four D7--branes (along with their mirrors)
be placed on each orientifold plane, resulting in an $\SO(8)^4$ non-abelian gauge
symmetry. This type IIB setup can also be obtained from the type I
string by a T-duality on both coordinates of the base \cite{Sen9605}.

Sen then argued that the F-theory moduli space close to one of the
orbifold fixed points, where $T^2 $ locally reduces to $\mathbb R^2$,
can be accurately described by the physics of the 4d $\Ncal =2 $, $\SU(2)$
Seiberg-Witten theory with four fundamental hypermultiplets. Moreover,
Banks \etal \cite{Banksetal96} showed that this gauge theory can be naturally
realised as the low energy effective theory on the worldvolume of a probe
D3--brane in the limit where the rest of the orientifold singularities are
taken to be very far away, and the  moduli space of the theory is captured
by the dynamics of the worldvolume fields.
By considering multiple coincident D3-branes as probes \cite{Douglasetal96,Aharonyetal96}
the $\SU(2)\cong \Sp(1)$ gauge group
can be extended to higher rank to obtain an $\Sp(N)$ gauge theory, at the
expense of introducing an extra hypermultiplet in the antisymmetric
representation of $\Sp(N)$.\footnote{The alternative extension
to $\SU(N)$ will be discussed in Section 5.
However, it is not the natural generalisation from the F-theory point of view.}

Let us summarise the setup and field content of the above physical-string configuration:
We will consider the low energy worldvolume action on a stack of $N$ coincident
D3--branes (and their mirrors) living in the $(x^0,\ldots,x^3)$ directions. These
probe the background generated by 4 D7s (and their mirrors) and a single O7--plane lying
in $(x^0,\ldots , x^7)$. The orientifold plane is added in such a way so
as to preserve the same 8 supersymmetries as the D3--D7 system and the
3-3 and 7-7 strings would  generate respective $\SU(2N)$ and
$\SU(8)$ gauge symmetries. However, since all the branes are sitting
at the  orientifold fixed plane, these project to $\Sp(N)$ and
$\SO(8)$ because of the orientation reversal action on the open string
Chan-Paton indices, which imposes  symmetric or antisymmetric
conditions on the gauge group matrices. Ramond-Ramond (RR) tadpole cancellation further
restricts one to only retain antisymmetric matrices
for the D7s; one is then forced to consider symmetric matrices for the
D3s \cite{GimonPolchinski96}.

In the low-energy limit,
the dynamical fields corresponding to 7-7 strings decouple and
$\SO(8)$ becomes a global symmetry of the system. The massless
 spectrum of 3-3 strings fluctuating in the worldvolume
$(x^0,\ldots ,x^3)$ and overall transverse $(x^8,x^9)$ directions
yields the degrees of freedom corresponding to the $\Ncal\! =\!2$ vector
multiplet in the adjoint (symmetric) representation of
$\Sp(N)$. The fluctuations in  the directions relatively transverse
to the D3s $(x^4,\ldots , x^7)$ furnish a hypermultiplet transforming
in the antisymmetric tensor representation of the gauge group, which
captures the motion of the D3s in these directions.
Therefore, the low
energy D3 worldvolume action describes 4d $\Ncal\! =\!2$ SYM with
gauge group $\Sp(N)$, four hypermultiplets in the fundamental and one
in the antisymmetric representation, sitting at the conformal
point of its moduli space.

\begin{table}[h]
\begin{center}
\begin{tabular}{|c|c|c|c|c|c|c|}
\hline
\textrm{Component} & $\SO(1,3)$ & $\SU(2)_a$ &$\SU(2)_A$ & $\Urm(1)_R$ & $\Sp(N)$ & $\mathrm{SO}(8)$ \\ \hline
$A,G$ & $(2,2)$ & 1 & 1 & $0$ & $ N(2N+1)$ &1 \\
$\phi$& $(1,1)$ & 1 & 1  & $+2$ & $ N(2N+1)$ & 1\\
$\phi^\dagger$& $(1,1)$& 1 & 1  & $-2$ & $ N(2N+1)$ & 1\\
$\lambda_{\alpha, a}$& $(2,1)$ & 2 & 1 & $+1$ &  $N(2N+1)$ & 1 \\
$\bar{\lambda}_{\dot{\alpha}, a}$& $(1,2)$ & 2 & 1 & $-1$ &  $N(2N+1)$ & 1 \\ \hline
$z_{aA}$& $(1,1)$& 2 & 2  & $0$ & $N(2N\!-\!1)\!-\!1\!$ & 1\\
$\zeta_{\alpha, A}$& $(2,1)$ & 1 & 2 & $-1$ &  $ N(2N\!-\!1)\!-1\!\!$ & 1 \\
$\bar{\zeta}_{\dot{\alpha}, A}$& $(1,2)$ & 1 & 2 & $+1$ &  $ N(2N\!-\!1)\!-1\!\!$ & 1 \\ \hline
$q_{a}^M$& $(1,1)$& 2 & 1  & $0$ & $ 2N$ & 8\\
$\eta_{\alpha}^M$& $(2,1)$ & 1 & 1 & $-1$ &  $2N$ & 8 \\
$\bar{\eta}_{\dot{\alpha}}^M$& $(1,2)$ & 1 & 1 & $+1$ &  $2N$ & 8 \\ \hline
\end{tabular}
\caption{The on-shell field content of the $N_f=4$ theory in component form.
The representations in the first column are actually in terms of the Euclidean
Lorentz group $\SO(4)\sim \SU(2)_L\times\SU(2)_R$.
The fundamental fields carry an $\SO(8)$ flavour index $M = 1,\ldots, 8$, while
the antisymmetric fields an $\SU(2)$ `flavour' index $A=1,2$. Note that $(z,\zeta,\bar{\zeta})$
transform in the irreducible second--rank antisymmetric representation of $\Sp(N)$, which
in the text we call ``antisymmetric'' for brevity. We write $\SO(8)$ rather than the more
accurate $\mathrm{O}(8)$ since we will not keep track of discrete groups.}
\label{tablesp}
\end{center}
\end{table}

As far as the global symmetries are concerned, the presence of the
D7--branes breaks the D3 transverse group of rotations down to
$\SO(4)\times \U(1)_R\subset \SO(6)$.
Furthermore, we write this $\SO(4)$ as $\SU(2)_a\times\SU(2)_A$,
$\SU(2)_A$ being a flavour-like symmetry for the antisymmetric
fields; no other field transforms nontrivially under its action. The
rest of the $\SO(6)$ global symmetry subgroup accounts for the
$\Ncal\! =\! 2 $ R-symmetry, $\U(2)_R\cong \SU(2)_a\times\U(1)_R$ and we remind
the reader that the fundamental
fields transform as vectors under the global $\SO(8)$ flavour group.
The precise transformation
properties of all degrees of freedom under the symmetries of the system
are summarised in Table \ref{tablesp}, which is adapted from \cite{Gavaetal99}.

By considering a large number of coincident D3--branes and taking their near--horizon
limit, it is possible to obtain the supergravity dual of the $N_f=4$ theory
in terms of strings in $\AdS_5\times
\mathrm S^5/\mathbb Z_2$, where $\mathbb Z_2$ is an orientifold action on the $\mathrm S^5$
\cite{FayyazuddinSpalinski98,Aharonyetal9806} (see also \cite{Ennesetal00}).
Instanton effects in the AdS/CFT context
have been studied in \cite{Gutperle99,Gavaetal99,Hollowood99}, while the plane--wave limit of
the theory has been investigated in \cite{Berensteinetal02,Gomisetal03}. Higher
derivative corrections were considered in \cite{Kochetal98,Kochetal99}, and
the geometry of the holographic dual of the Higgs branch of the theory was described in
\cite{Guralniketal04}. Recently, \cite{KomargodskiRazamat07} used the AdS/CFT dual
to discuss the behaviour of \emph{strongly coupled} $N_f=4$ scattering amplitudes.

\subsection{The spacetime action}\label{spacetime}

We now turn to the construction of a Lagrangean for the above $\Ncal=2$ theory by
taking its formulation in terms of $\Ncal=1$ superfields as a starting point.\footnote{Our
notation and conventions are summarised in Appendix \ref{notation}. General
reviews of superspace techniques and $\Ncal=2$ supersymmetric gauge theories can
be found, for instance, in \cite{Superspace,Sohnius85,AlvarezGaumeHassan97,Kovacs99,Bilal01}.}
This reads
\be\label{N2-4}
\begin{split}
 \Lcal  = & \frac{1}{8\pi}\mathrm{Im}\; \mathrm{Tr}\left[ \tau  \left( \int
   d^2\theta\;  W^\alpha  W_{\alpha} + 2
  \int d^2 \theta d^2\bar{\theta}\; e^{2V}\Phi^{\dagger}e^{-2
   V} \Phi \right)\right]+
  \int d^2 \theta d^2\bar{\theta}\; Q^{\dagger I} e^{-2 V}Q_I \\
& +   \int d^2 \theta d^2\bar{\theta}\; Q'^I e^{2 V}Q'^\dagger_I +
\mathrm{Tr}\left(    \int d^2 \theta d^2\bar{\theta}\;
  e^{2V}Z^{\dagger}e^{-2 V} Z +   \int d^2 \theta d^2\bar{\theta}\;
   e^{-2V}Z^{\prime} e^{2 V} Z'^\dagger\right)\\
& +   \sqrt{2} \left(\int d^2\theta(Q'^I  \Phi Q_I + \mathrm{Tr}\left(
   Z' [ \Phi,  Z ]\right)  ) + h.c. \right) \; .
\end{split}
\ee
 The $\Ncal=2$ vector multiplet consists of the $\Ncal=1$ vector and chiral superfields
$( V,\Phi)$, the antisymmetric hypermultiplet of the chiral
and antichiral $(Z,Z'^\dagger)$ and the four
fundamental hypermultiplets of the four chiral and four antichiral superfields
$(Q^I,Q'^{\dagger I})$ respectively. $(Q^{\dagger I} , Q'^I)$ are four antichiral
and chiral superfields
transforming in the conjugate fundamental representation and the $\SU(4)$ flavour
index $I$ runs from 1 to 4. The fundamental representation of $\Sp(N)$ is
pseudoreal, which means that it is related to its conjugate simply by raising
and lowering indices. The flavour symmetry is thus enhanced
to $\SO(8)$. However in this $\Ncal = 1$ notation
this $\SO(8)$ flavour symmetry is not explicit.
It is instead implicitly realised via the subgroup  $\textrm{SU}(4)\times \textrm{U}(1)\subset
\textrm{SO}(8)$ and the decomposition $\mathbf{8_{\mathrm{s}}= 4_1
 +\bar 4_{-1}}$, which reflects the fact that we are considering four kinds of 3--7
and 7--3 strings.
Also hidden in (\ref{N2-4}) is the $\SU(2)_A$ symmetry, which we will restore
in due course together with explicit $\SO(8)$ invariance. Lastly, the
$\SU(2)_a$ part of the $\Ncal=2$ R--symmetry is also not manifest at this stage.
The complexified coupling
is $\tau =\frac{\Theta_{YM}}{2\pi}+\frac{4\pi i}{g^2} $ but, since we are
only interested in the perturbative behaviour of the theory, we can
safely set the total derivative terms to zero by requiring that
$\Theta_{YM}=0$. We will also ignore any total derivative terms coming
from integration by parts.

In component form we
have $(A^\mu, \lambda, {\sf D})$ for $ V$, $(\phi, \chi, {\sf F}_\phi)$
for $\Phi$ and  $(q,\eta,{\sf F}_q)$ for $Q$, with similar superfield expansions for
$Q'^\dagger$, $Z$ and $Z'^\dagger$.
Since we are
constructing this $\Ncal=2$ action out of $\Ncal=1$ quantities, the
coupling appearing in front of the
superpotential terms can, in principle, be different to the coupling
of the kinetic terms for the $\Ncal=2$ vector multiplet. However, $\Ncal=2$
supersymmetry requires that they all be equal \cite{Sohnius85}. After expanding the superfields and performing
the Grassmann integration one obtains the expression
\be\label{neq2-4}
\begin{split}
\Lcal  = &\frac{1}{g^2} \mathrm{Tr} \left( -\frac{1}{4}  F^2 +
 (D^\mu \phi)^{\dagger} (D_\mu \phi) - i \lambda\; \Dslash \bar{\lambda} - i
\bar\chi \Dslash \chi   - i
 \sqrt{2}\; [\lambda,\chi] \phi^\dagger - i \sqrt{2}\; [\bar{\lambda},
 \bar{\chi}] \phi
 \right)\\
&  +
(D^\mu q)^{\dagger I}(D_\mu q)_I +(D^\mu q')^I(D_\mu
 q')_I^{\dagger}  - i \bar \eta^I \Dslash
{\eta}_{I}  - i
\eta'^I \Dslash \bar{\eta}'_{I}-
i \sqrt 2 \; q^{\dagger I} \lambda \eta_{I}\\
&  + i\sqrt{2}\;
   \bar{\eta}^I\bar{\lambda} q_I  - i\sqrt{2}\;
  q'^I \bar \lambda \bar{\eta}'_{ I} + i\sqrt{2}\;
   \eta'^I \lambda q'^\dagger_I   + \mathrm{Tr} \left(
(D^\mu z)^{\dagger}(D_\mu z) + (D^\mu z')(D_\mu z')^\dagger\right.\\
& - i \bar \zeta \Dslash
\zeta   - i
 \zeta' \Dslash \bar \zeta'
 - i\sqrt{2}\; [\lambda, \zeta]  z^{\dagger} \left. - i\sqrt{2}\;
   [ \bar{\lambda},\bar \zeta ]z  - i\sqrt{2}\;  [\bar\lambda, \bar\zeta']z'  \left. - i\sqrt{2}\;
       [\lambda, \zeta'] z'^\dagger \right)\right. \\
&  - \sqrt 2 \left[\left( \eta'^I\chi q_I +
\eta'^I  \phi \eta_{I}   +   q'^I \chi
\eta_{I} \right)+  \mathrm{Tr} \left(-[\chi, \zeta'] z+ \zeta' [\phi, \zeta]+ [\chi ,\zeta]z'\right) + h.c.\right]- V_S \; ,
\end{split}
\ee
where our convention for the covariant derivative is
$D_{\mu} = \partial_{\mu} -i A^a_{\mu}T^a_{R} $ and
$\Dslash^{\dot\alpha \alpha} = (\bar\sigma^{\dot\alpha\alpha})^\mu D_\mu$.
$V_S$ is the scalar potential obtained by integrating out the ${\sf F}$- and ${\sf D}$-terms. It is given by
\be
V_S = {\sf F}^\dagger_{q }{\sf F}_{q} + {\sf F}_{ q'}{\sf F}^\dagger_{q'} +
\mathrm{Tr}\left( {\sf F}^\dagger_{z}{\sf F}_{z}+ {\sf F}_{z'}{\sf F}^\dagger_{z'}+
\frac{1}{g^2}{\sf F}^\dagger_{\phi}{\sf F}_{\phi}\right)+\frac{1}{2g^2}{\sf D}^2\; ,
\ee
where the individual terms with their index
structure made explicit are
\bea
\nn ({\sf F}_{q})_I^{ i} =-\sqrt 2\;
(\phi^\dagger)_{\phantom i j}^{i} q_I^{' \dagger j}&,& ({\sf F}_{
q'})^I_i = -\sqrt 2 \; q_{j}^{\dagger I } (\phi^\dagger)_{\phantom j
i}^j\\  ({\sf F}_{z})^i_{\phantom i j} = -\sqrt 2 \left[
\phi^\dagger,z'^\dagger\right] ^i_{\phantom i j} &,&
({\sf F}_{z'})^i_{\phantom i j} = -\sqrt 2 \left[
z^\dagger , \phi^\dagger \right]^i_{\phantom i j}
\eea
and
\bea
 \nn
 ({\sf F}_{\phi})^i_{\phantom i j} &=& - g^2\sqrt 2  \left[z'^\dagger,z^{\dagger
  }\right]^i_{\phantom i j}-  \frac{g^2}{\sqrt 2 } \left(
q'^{\dagger i}_I q_j^{\dagger I}+   q'^\dagger_{ I j}
q^{\dagger I i} \right)\\
{\sf D}^a &=& - \mathrm{Tr} \left(T^a[ \phi^\dagger , \phi]+
g^2 T^a[ z^\dagger , z] - g^2 T^a[ z',z'^\dagger ] \right)+g^2\left( q^{ \dagger
  I} T^a q_I -  q'^I T^a  q'^\dagger_I \right)\; .
 \eea
The $(T^a)_{\phantom i j}^i$'s are the generators of the
fundamental representation of $\Sp(N)$ and in obtaining the full scalar
potential one also needs to make use of the following identity
\be
\label{fierz}
(T^a)^i_{\phantom i j} (T_a)^k_{\phantom k l} = \frac{1}{2}(
 \delta^i_l\delta^k_j - \Omega^{i
  k}\Omega_{j l} )\; .
\ee

To further reorganise the action (\ref{neq2-4}), we recall that the twistor approach to
gauge theory amplitudes breaks the symmetry between positive and negative helicity
states \cite{Witten0312}. Here
we implement this by splitting the action into a piece independent of the gauge coupling and
another piece which is of order $g^2$. This is done by performing a series of rescalings which
read as follows: For the adjoint fields we have
\be\label{read}
 (\phi,\phi^\dagger) \ra  (ig\sqrt 2\phi,-\frac{ig}{\sqrt 2} \phi^\dagger) \quad , \quad (\lambda,\bar\lambda,) \ra
(g^{1/2}\lambda,g^{3/2}\bar\lambda,) \quad , \quad
(\chi,\bar\chi) \ra  (g^{1/2}\chi,g^{3/2}\bar\chi)\;,
\ee
for the antisymmetric ones
\bea\label{reantisym}
\nn (z, z^\dagger)\ra  (z, z^\dagger) &,&
\nn (z', z'^\dagger)\ra (i z', -iz'^\dagger)\\
(\zeta,\bar{\zeta}) \ra
(-\frac{i\zeta}{ g^{1/2}\sqrt 2 },i g^{1/2}\sqrt 2 \bar{\zeta}) &,&
 (\zeta',\bar{\zeta'}) \ra
(\frac{\zeta'}{g^{1/2}\sqrt 2 },g^{1/2}\sqrt 2 \bar{\zeta}')\, ,
\eea
while for the fundamentals
\bea\label{refun}
\nn (q_I, q^{\dagger I})\ra  (q_I, q^{\dagger I}) &,&
\nn ({q'}^I,{q'}_I^\dagger)\ra  ( i q'^I, -i q'^\dagger_I)\\
 (\eta_{ I} ,\bar{\eta}^I) \ra
(-\frac{i\eta_I }{g^{1/2}\sqrt 2 },i g^{1/2}\sqrt 2 \bar{\eta}^I) &,& (\eta'^I,\bar{\eta}'_{I}) \ra
(\frac{\eta'^I}{g^{1/2}\sqrt 2 },g^{1/2}\sqrt 2\bar{\eta}'_{I})\:.
\eea

We will also make  the symmetries of Table \ref{tablesp} explicit by appropriately arranging
the antisymmetric fields into $\SU(2)_A$ doublets and collecting the fundamentals into
$\SO(8)$ spinors (which can be exchanged for vectors by $\SO(8)$ triality). We finally collect
the hypermultiplet scalars and the adjoint fermions into doublets of
$\SU(2)_a$. The above statements are summarised by the definitions
\bea\label{doublets}
\nn \lambda^a = \doublet{\lambda}{-\chi} \quad , \quad \bar \lambda_a =
\left( \bar \lambda,-\bar \chi\right) &,&
 \bar\lambda^a = \doublet{-\bar\chi}{-\bar\lambda} \quad , \quad \lambda_a =
\left( \chi,\lambda \right)\\
\nn \eta^M =
 \doublet{\eta_I}{\eta'^I}  \quad , \quad  \bar \eta_M = \left( \bar\eta^I , \bar \eta'_I \right)&,& \bar\eta^M = \doublet{\bar
 \eta'_I} { \bar\eta^I} \quad , \quad
\eta_M = \left( \eta'^I , \eta_I\right)\\
\nn \bar \zeta^A = \doublet{\bar\zeta}{\bar\zeta'} \quad , \quad \zeta_A =
\left( \zeta,\zeta'\right) &,&
 \zeta^A = \doublet{\zeta'}{-\zeta} \quad , \quad  \bar\zeta_A =
\left( -\bar\zeta',\bar\zeta\right)\\
\nn   z^a_{\phantom a A} =\twobytwo{z}{z'}{-z'^\dagger}{z^\dagger } &,&
 z^A_{\phantom A a} =
 \twobytwo{z^\dagger}{-z'}{z'^\dagger}{z} \\
  q^{ a}_{\phantom a M} = \twobytwo{-q'^I}{q'^\dagger_I}{-q^{\dagger I}}{q_I} &,&
 q^M_{\phantom M a} =
 \twobytwo{-q'^\dagger_I}{-q_I}{-q^{\dagger I}}{-q'^I}\;.
 \eea
Having made the $\SO(8)$ flavour symmetry of the fundamental fields manifest in terms
of components, we can also  collect them into 8 $\Ncal=2$
 `half-hypermultiplets' $Q_M=(\bar{\eta}_M,q^a_M,\eta_M)$,
each of which contains two bosonic and two fermionic fields. This type of
multiplet arises only for pseudoreal representations, allowing a description
in terms of half the usual field content of $\Ncal=2$ supersymmetry
\cite{BreitenlohnerSohnius81,MezincescuYao84,Sohnius85}. Note that it is not
possible to have a description in terms of full $\Ncal=2$ hypermultiplets that
manifestly preserves the $\SO(8)$.

For the gauge field we introduce an anti-selfdual two-form $G^{\mu\nu}$ as
 a Lagrange multiplier, via which (up to a topological term which will not play a role in
our perturbative study) we can rewrite the Yang--Mills action in first order form \cite{Siegel99}
 \be
-\frac{1}{4g^2}\Tr F^\mn F_\mn\rightarrow -\frac{1}{2}\mathrm{Tr} \left(G_\mn F^\mn-\frac{1}{2}g^2 G_\mn G^\mn \right)\;.
\ee

The final expression for the action, including the full quartic
contributions arising from the scalar potential, takes the form
\be\label{Nf4full}
\begin{split}
\mathcal L =  & \mathrm{Tr} \left[ -\frac{1}{2}  GF +\frac{1}{4}g^2G^2 +
  D\phi^{\dagger} D\phi + i \bar \lambda^a \Dslash \lambda_a
 -   \lambda^a\lambda_a \phi^{\dagger} + 2 g^2
 \bar\lambda^a\bar\lambda_a  \phi  \right]  + \mathrm{Tr} \left[\frac{1}{2}
  D z^a_{\phantom a A}D z^A_{\phantom A a}\right.\\
&\left.      + i \bar\zeta^A  \Dslash
\zeta_A  -  z^a_{\phantom a A}[\lambda_a,
 \zeta^A] - 2 g^2 z^A _{\phantom A a}[\bar\zeta_A , \bar \lambda^a] +
 \zeta^A\zeta_A \phi - 2  g^2 \bar\zeta^A \bar\zeta_A
 \phi^\dagger\right]  + \frac{1}{2}  Dq^{ a}_{\phantom{ a}M}  Dq_{\phantom M a}^M\\
&  - i \bar \eta_M \Dslash
\eta^M    +  q^{  a}_{\phantom{ a}M} \lambda_a \eta^M  - \frac{1}{2}\eta_M\phi\eta^M -2 g^2\left(
   \bar\eta_M \bar{\lambda}^a q^M_{\phantom M  a }   +\frac{1}{2} \bar \eta_M \phi^\dagger\bar\eta^M\right)\\
&+ g^2 \left(-\frac{1}{2} q^a_{\phantom a M}
  \{\phi^\dagger , \phi \} q^M_{\phantom M a} +\frac{1}{4}  q^{
 a}_{\phantom a M}
  [z^b_{\phantom b A},z^A_{\phantom A a}] q^M_{\phantom M b}  \right)
 - \frac{g^2}{8}\left( (q^{ a}_{\phantom a M} q^N_{\phantom N a})(q^{
 b}_{\phantom b N} q^M_{\phantom M b})\right.\\
 & \left.
  + (q^{ a }_{\phantom a M} q^{ b}_{\phantom b N} )(q^N_{\phantom N a}
 q^M_{\phantom M b}) \right)- g^2\;   \mathrm{Tr}  \left( \frac{1}{2}[\phi^\dagger, \phi]^2
  + \frac{1}{4} [ z^a_{\phantom a A},z^A_{\phantom A b } ] [z^b_{\phantom b
  B},z^B_{\phantom B a}] + [z^a_{\phantom a A } , \phi][
 \phi^\dagger ,z^A_{\phantom A a}] \right) \; .
\end{split}
\ee
By taking the $g\rightarrow 0$ limit one obtains the `selfdual' truncation of the Lagrangean,
which has the same field content but only a subset of the interactions of the full theory.
The $\mathcal O(g^2)$ terms can be thought of as perturbations around the selfdual theory.

 In anticipation of the twistor approach, we will perhaps surprise the
reader by once again hiding the global $\SO(8)$ symmetry that we just made manifest.
This is done by decomposing the flavour index $M\rightarrow A'\otimes X$
according to the special maximal embedding $\SO(8)\supset\SU(2)_{A'}\times\Sp(2)$ where
the indices run over $A'=1,2$ and $X = 1,\ldots , 4$. One motivation for this is
that each doublet indexed by $A'$ has the field content of a full $\Ncal=2$ hypermultiplet,
but the main reasoning behind it will become clear in the next section. In the interim---and
to facilitate comparison with the twistor analysis---we include the action
for this selfdual truncation, which takes the simple form
\bea\label{selfdualsp}
\mathcal{L} &=&   \mathrm{Tr} \left[ -\frac{1}{2}  GF  +
   D\phi^{\dagger} D\phi + i \bar \lambda^a \Dslash \lambda_a
 -    \lambda^a\lambda_a \phi^{\dagger}   \right] \\ \nn
&&\!\!\!\!     - \mathrm{Tr}\! \left[\frac{1}{2}
   D z^{aA}D z_{aA}  + i \bar\zeta^A  \Dslash
\zeta_A  +  z^{aA}[\lambda_a,
 \zeta_A] + \zeta^A\zeta_A \phi \right]\\ \nn
&&\!\!\!\!  -\left( \frac{1}{2}  Dq_{a A' X} Dq^{ a A' X} + i \bar \eta_{A' X}\Dslash
\eta^{A' X}   +
   q_{a A' X} \lambda^a \eta^{A' X}  + \frac{1}{2}\eta_{A' X}\phi\eta^{A' X}\right)
 \; .
\eea
This action is the $N_f=4$ analogue of the selfdual truncation of $\Ncal=4$
SYM introduced by Siegel \cite{Siegel9205}.

\section{Twistor strings} \label{Twistorsec}
We now turn our attention to constructing a twistor string dual to the gauge theory we have
just described. To begin with, we will briefly review the most relevant
parts of  the twistor string dual for $\Ncal =4$, $\SU(N)$
gauge theory in four dimensions \cite{Witten0312}\footnote{To be precise, Witten studied the $\U(N)$ theory
but since that only involved gluon amplitudes it is essentially the same to consider $\SU(N)$;
the gluons don't couple to the $\U(1)$ `photon'. Moreover, when considering colour-stripped amplitudes the
extra $\U(1)$ piece will not affect the results, even for external
scalars or fermions.}, subsequently modifying it appropriately
for the $N_f = 4$ case.

\subsection{The open B--model}

In this section we collect a few well--known facts on the open topological
B--model which will be useful in what follows. This is not intended to be
a thorough review, for which we refer the reader to \eg
\cite{NeitzkeVafa0410,Marino0410,Vonk05,MirrorSymmetry}.
The B--model \cite{Witten91} arises as an axial--type twisting of the $\Ncal=(2,2)$
supersymmetric 2d nonlinear $\sigma$--model, which turns out to
only be consistent when the target space is Calabi--Yau. For a bosonic target space,
the worldsheet field content of the theory consists of bosonic scalars $\phi^m,\phi^{\bar{m}}$
providing the map to the target manifold, and ghost--number one fermions
$\eta^{\bar{m}}$ and $\theta_m$ (plus a worldsheet one--form $\rho^m_{\bar{z}}$ which
will not play a role in our analysis).
The action of the BRST charge $Q_B$ on these fields is such that it
can be precisely mapped to the Dolbeault operator $\dbar$ on the target space
Calabi--Yau, and this identification leads to the following well--known relations
between worldsheet fields and the geometry of the target space
\be \label{mapping}
\phi^m \sim Z^m\;,\; \phi^{\bar{m}}\sim \bar{Z}^{\bar{m}}\;,\;
\eta^{\bar{m}} \sim  d\bar{Z}^{\bar{m}}\;,\;
\theta_m \sim  \frac{\partial}{\partial Z^m} \;.
\ee
We will only be interested in the BRST transformations of these fields in
the presence of a boundary, which are given by \cite{Witten9207}
\be
\begin{split}
\delta_B \phi^m=0\;,\quad \delta_B \phi^{\bar{m}}=i\alpha \eta^{\bar{m}}\;,\quad
\delta_B\eta^{\bar{m}}=0\;,\quad \delta_B\theta_m=0\;.
\end{split}
\ee
This implies (see \eg \cite{Brunneretal99,KatzSharpe02}) that imposing Neumann boundary
conditions along a particular holomorphic direction (say $m$) requires that $\theta_m=0$,
while imposing Dirichlet directions along an antiholomorphic direction $\bar{m}$ leads to
$\eta^{\bar{m}}=0$.

A generic open string vertex operator, giving rise to a local observable, can be written as
\be
\Vcal=
\theta_{m_1}\cdots\theta_{m_p}\eta^{\bar{n}_1}\cdots\eta^{\bar{n}_q}
{{V(\phi,\bar{\phi})^i_{\;\;j}}^{m_1\cdots m_p}}_{\bar{n}_1\cdots\bar{n}_q}
\ee
where $i$,$j$ denote the Chan--Paton indices. BRST invariance of this operator
requires that $V$ be a  $(0,q)$--form with values in
$\wedge^p T^{(1,0)}$ (times the Chan--Paton group).
Since \emph{physical} open string vertex operators arise at ghost number one, in practice
one needs to consider two types of vertex operators
\be \label{VO}
\text{(a)}\;\; \Vcal=\eta^{\bar{m}} V^i_{\;\;j\bar{m}}\qquad
\text{and}\qquad\text{(b)}\;\; \Vcal'=\theta_m {V^{\prime i}_{\;\;\;j}}^m\;.
\ee
Recalling the identifications in (\ref{mapping}), we see that these states correspond
to either matrix--valued $(0,1)$--forms or tangent vectors on the target manifold.
Therefore, when considering space--filling (`D5') branes
on the Calabi--Yau \cite{Witten9207}, by imposing Neumann--Neumann (NN)
boundary conditions on all open strings, the physical open string spectrum is just given by
a $(0,1)$ form $\Acal=\diff \bar{Z}^{\bar{m}} \Acal_{\bar{m}}$. The target space
interactions can be encoded in the cubic holomorphic Chern--Simons theory
\be \label{HCSorig}
S=\frac{1}{2}\int_{\mathrm{CY}}\mathbf{\Omega}\wedge\textrm{Tr}
(\Acal\cdot\bar{\partial}\Acal+\frac{2}{3}\Acal\wedge\Acal\wedge\Acal)\,
\ee
which is written with the help of the $(3,0)$ holomorphic volume form of the Calabi--Yau.

In the following we will assume the straightforward generalisation of the above
statements to the super--Calabi--Yau case.

\subsection{Review of the dual for $\Ncal=4$ SYM}

In \cite{Witten0312}, Witten showed that the tree level $n$-gluon MHV amplitudes, that is
the amplitudes with $n-2$  positive and $2$ negative helicity
gluons (when all external particles are taken to be outgoing) can be
reconstructed from an open string theory in supertwistor space.
Essential to this was the observation that
these amplitudes localise on holomorphically embedded, degree--one
curves of genus zero in $\Supertwistor$, and the string theory in question is the open
string sector of the topological B--model with  $\Supertwistor$  target
space, which is well defined since the latter is
a super-Calabi-Yau. The isometries of $\Supertwistor$ encode the
$\PSU(2,2|4)$ superconformal symmetry of the $\Ncal=4$ theory in a
linear way, while the open string sector is realised by
introducing Euclidean `D5'-branes
wrapping the bosonic directions of $\Supertwistor$ $(Z,\bar Z)$ but only the
holomorphic part of  the fermionic directions
$\psi^I$ ($I=1,\ldots,4$).
This can be interpreted as a localisation
of the D5s in the transverse fermionic coordinates and in \cite{Witten0312} this
locus was taken to be at $\bar \psi^{\bar I} =0$.
Since this imposes Dirichlet boundary conditions only on the
antiholomorphic fermionic directions $\bar{\psi}^{\bar{I}}$ (which would not have been
possible had they been bosonic), it follows that $\theta_{m},\theta_{I}=0$ and
from (\ref{VO}) we see that the only physical field
is a nonabelian $(0,1)$--form $\Acal= d\bar{Z}^{\bar{m}}\Acal(Z,\bar{Z},\psi)_{\bar{m}}$,
which in addition is independent of the antiholomorphic fermionic coordinates.
Therefore the superfield expansion of $\Acal$ is
\be \label{N=4 superfield}
\Acal=A+\psi^I\lambda_I+\frac{1}{2!}\psi^I\psi^J\phi_{IJ}+
\frac{1}{3!}\epsilon_{IJKL}\psi^I\psi^J\psi^K\tilde{\lambda}^L+
\frac{1}{4!}\epsilon_{IJKL}\psi^I\psi^J\psi^K\psi^LG\ ,
\ee
where we will from now on suppress the gauge indices and form structure.

As mentioned above, the open string field theory of the
B-model  reduces to  a holomorphic version of Chern-Simons theory
\cite{Witten9207}, which can be straightforwardly extended to
super--Calabi-Yau manifolds, yielding the following action \cite{Witten0312}
\be \label{HCS}
S=\frac{1}{2}\int_{\mathrm{D5}}{\mathbf \Omega}\wedge\textrm{Tr}
(\Acal\cdot\bar{\partial}\Acal+\frac{2}{3}\Acal\wedge\Acal\wedge\Acal)\
, \ee
where in this case
\be\label{volume}
  {\mathbf \Omega} ={\frac{1}{4!}{\mathbf \Omega'}\epsilon_{IJKL}
d\psi^Id\psi^Jd\psi^Kd\psi^L }\; \qquad \left(\text{with} \quad
{\mathbf \Omega'}={\frac{1}{4!}\epsilon_{IJKL}Z^I dZ^JdZ^KdZ^L}\right)\;,
\ee
is the globally defined holomorphic volume form.\footnote{Note that, as mentioned in
\cite{Witten0312}, ${\mathbf \Omega}$ does not actually define a top form in the
fermionic directions and ideally should be promoted to a so--called \emph{integral}
form, which does. A thorough discussion of integration on supermanifolds in
similar contexts appears in \cite{GrassiPolicastro04}, where
more references can be found. However, as in \cite{Witten0312}, the choice of $\mathbf\Omega$ in
(\ref{volume}) appears to be sufficient for our purposes, and we will content ourselves
with this na\"ive choice in the following.}
The classical equations of motion following from (\ref{HCS}) are
\be
\dbar\Acal +\Acal\wedge\Acal=0 \;,
\ee
while an infinitesimal gauge transformation takes the form\footnote{Here and in the
following we  use the
standard commutator of forms $[\alpha_p,\beta_q]=\alpha_p\wedge\beta_q-(-1)^{pq}\beta_q
\wedge\alpha_p$.}
\be
\label{Agt}
\delta\Acal  =  \dbar\epsilon +[\Acal , \epsilon]\;.
\ee
By  linearising around the trivial solution $\Acal = 0$, the above reduce to
$\dbar\Acal =0$ and $\Acal ' - \Acal  = \dbar\epsilon$ respectively, which
show that $\Acal$ is in the $\bar{\partial}$ cohomology class ${\mathrm H}^1$ and
thus a good physical state of the open B-model.
As is further explained in \cite{Witten0312}, the $\psi^I$s  carry an additional $\U(1)_S$ $(+1)$
charge, under which the B--model is anomalous and the superfield $\Acal$ is neutral.
The component fields $(A,\lambda_I,\phi_{IJ},\tilde{\lambda}^I,G)$
carry charge $(0,-1,-2,-3,-4)$ under this symmetry
and are then $(0,1)$--forms with values in the line bundle
$\Ocal (-k)$, where $-k$ is the appropriate S--charge. Each component field in
(\ref{N=4 superfield}) is then an element of
the sheaf cohomology class $\mathrm H^1(\Twistorspace , \Ocal (-k))$\footnote{Actually, as
noted in \cite{Witten0312}, the class is really $\mathrm H^{1}({\Twistorspace}', \Ocal (-k))$ where ${\Twistorspace}'$ is
a suitable open set of supertwistor space. However, we will ignore such subtleties here.} and
the Penrose transform \cite{Penrose67} maps these fields to the space
of solutions of massless free wave equations for fields of helicity
$1-k/2$ in Minkowski space. In this fashion one  recovers the spectrum of  $\Ncal =4 $ SYM.

The action (\ref{HCS}) thus contains all the fields of $\Ncal = 4$ SYM and at
least some of the interactions.
It does not contain \emph{all} the interactions, however. Rather, it describes the
subset corresponding to the so-called
selfdual sector of $\Ncal = 4$ SYM, as can be seen via a nonlinear form
of the Penrose transform, which precisely maps the hCS
action to this selfdual truncation \cite{PopovSamann04}. In order to recover the full set of
interactions it is necessary  to introduce nonperturbative objects, called
`D1-instantons' in \cite{Witten0312}, which are Euclidean 2-branes wrapping
the curves on which the desired amplitudes are supported. We will postpone
a review of these aspects to Section \ref{N4review}, and concentrate for
the moment on obtaining the analogue of the above construction for the
$N_f=4$ theory.

\subsection{Orientifolding the twistor string} \label{orientsec}

Having reviewed how the spectrum and selfdual interactions of $\Ncal=4$ SYM can be
recovered from the B--model on $\Supertwistor$, we now begin the analogous
construction for the $N_f=4$ theory. It is clear
from the above that the problem can be split into two  steps:
First, we will need to recover a B--model target space action  corresponding to
the selfdual sector of the gauge theory, and then, introducing D1--instantons
wrapping appropriate curves, we can proceed to reproduce the non-selfdual
amplitudes of the theory. In the following we will focus on the former part,
while the second step will be considered in Section \ref{Amplitudes}.

Following the intuition gained from the IIB description of the $N_f=4$ theory,
reviewed in Section \ref{ftheorysec}, and the
twistor description of quiver gauge theories in \cite{ParkRey04, Giombietal04}, it
is clear that some
sort of fermionic orientifold projection will be necessary in our approach.\footnote{Orientifolds
in a topological string context were first considered (for the A-model) in \cite{SinhaVafa00}.}
 We begin by considering the $\Ncal=4$ setup
of the previous section, choosing the number of `D5' branes to be $2N$. This produces
an $\SU(2N)$ gauge group, and accordingly the indices of $\Acal^i_{\;\; j}$ run
over  $i,j=1,\ldots , 2N$. Conformal invariance of the dual gauge theory requires
us to choose the orientifold action such as to leave the bosonic part of $\Supertwistor$
fixed.\footnote{Although we should emphasise that, in order to discuss a specific spacetime
signature, we will eventually need to pick a contour (e.g. ${\mathbb R} {\mathrm P}^3$)
within $\Twistorspace$, which can be imposed via a bosonic orientifold-type
operation (albeit a trivial one from our perspective, being already present for
$\Ncal=4$ \cite{Witten0312}). We thank Dave Skinner for a relevant discussion.}
However, we would like to reduce the amount of supersymmetry, which implies that the
orientifold should act on the fermions $\psi^I$ asymmetrically, in order to break the
$\SU(4)_R$ symmetry. Therefore, we begin by splitting the four fermionic coordinates
$\psi^I$ of $\Supertwistor$ into $I=\{a,A\}$, with $a=1,2$ and $A=3,4$.
The appropriate orientifold action is the combination of a $\mathbb Z_2$
orbifold (acting trivially on the Chan--Paton indices), the worldsheet parity
transformation $\hat{\omega}$ and an action on the Chan-Paton
indices brought about by acting with an antisymmetric hermitian matrix
$\tilde{\gamma}=i\Omega$, where
$\Omega_{2N\times2N}$ is the $\Sp(N)$ invariant tensor (see appendix \ref{notation})
\be\label{superorientifold}
\begin{split}
(a)&\quad  \psi^a\ra \psi^a \quad,\quad \psi^A\ra -\psi^A \\
(b)&\quad \Acal^i_{\;\;j}\ra \Omega^{ik}(\Acal^T)_k^{\;\;l}\Omega_{lj}
=(\Acal^T)^i_{\;\;j}
\equiv \Acal_j^{\;\;i}\;,
\end{split}
\ee
which is a superorientifold operation in $\Supertwistor$.\footnote{In writing
$(b)$ we have assumed that, as in the physical string case \cite{GimonPolchinski96},
$\hat{\omega}$ has eigenvalue $-1$ on
the $(0,1)$--form vertex operator $\Acal$. This minus combines with the $i^2$ from
$\tilde\gamma=i\Omega$ to give an overall plus in $(b)$.} Note that
the orbifold action $(a)$ breaks the fermionic coordinate symmetry
$\SU(4)_R\rightarrow\SU(2)_a\times\SU(2)_A$.\footnote{We choose the subscripts
having in mind the eventual identification of these symmetries with their
spacetime counterparts. }
Also note that it leaves the holomorphic volume form (\ref{volume}) invariant,
indicating that the target space is still super--CY and that we can
legitimately define a proper B--model action. In $(b)$ we have used $\Omega$
to raise and lower indices.

Requiring $\Acal$ to be invariant under this operation (which, on lowering
indices, translates to $\Acal_{ij}=\Acal_{ji}$), and considering its action on
the various component fields in the expansion (\ref{N=4 superfield}),
it is easy to see that one obtains the following decomposition
\bea\label{VZ}
\nn\hat \Acal &=&(A+\psi^a\lambda_a+\psi^1\psi^2\phi+\psi^3\psi^4\phi^\dagger
+\epsilon_{cd} \psi^3\psi^4\psi^c\tilde{\lambda}^d+
\psi^1\psi^2\psi^3\psi^4 G)\\
&&+\; \psi^A(
\zeta_A+\psi^a z_{Aa}+\epsilon_{AB}\psi^1\psi^2\tilde{\zeta}^B)\\
\nn &=&\mathcal V + \psi^AZ_A\\
\nn &=&\mathcal V + \mathcal Z\;,
\eea
where in the first line we have collected the terms ($\mathcal V$)
which are symmetric (when both indices are either up or down) under the orientifold
operation of (\ref{superorientifold}). Since these have $N(2N+1)$ gauge degrees of freedom,
we immediately conclude that they transform in the adjoint
representation of $\Sp(N)$. Similarly in the second line we have
displayed the terms ($\mathcal Z$) which are antisymmetric under said operation
and therefore have $N(2N-1)-1$ degrees of freedom and transform in the (second--rank)
antisymmetric tensor representation of $\Sp(N)$.

By repeating the analysis performed for the $\Ncal =4$ theory and
studying the linearised classical equations of motion around
the trivial solution $\hat \Acal=0$, one obtains the superorientifold-invariant
elements of the (Dolbeault) cohomology, which via the Penrose transform map to
part of the spectrum of the $N_f = 4$ theory \cite{Witten0312}. In a helicity basis this is
\bea\label{VZspectrum}
\nn\mathcal V \quad=\underbrace{(A,\lambda_a,\{\phi,\phi^\dagger\},\tilde{\lambda}^a ,G)}_{\textrm{\small{1--forms of S-charge $(-k)$ in twistor space}}}&\stackrel{\textrm{\small{Penrose}}}{\longleftrightarrow}&\underbrace{(A,\lambda_a,\{\phi,\phi^\dagger\},\bar{\lambda}^a ,G)}_{\textrm{\small{fields of helicity $(1-k/2)$ in Mink. space}}} \\
 \mathcal Z \quad= \;\;\qquad\qquad\overbrace{(0,\zeta_A,z_{aB},\tilde{\zeta}^A,0)} \qquad \;\;\;\qquad&\stackrel{\textrm{\small{Penrose}}}{\longleftrightarrow}&\;\;\;\qquad\qquad \overbrace{(0,\zeta_A,z_{aB},\bar{\zeta}^A,0)}
\eea
and we have, therefore,  obtained the adjoint and antisymmetric sector of the $N_f=4$ theory.
However, to complete the derivation of the spectrum on the twistor side, we still
need to recover the fundamental degrees of freedom, to which  we now turn our
attention.

\subsection{Flavour-branes and the Fundamental Sector} \label{fundsec}

By analogy with the IIB string description, it should be clear that incorporating the
fundamental fields of the $N_f=4$ theory
will require the introduction of a new object in twistor space.
We will implement this by adding a new kind of brane to our configuration, which we will
call a `flavour'-brane, as it roughly corresponds to a D7--brane in
the physical string setup, in the sense that strings stretching between the
`D5's and the flavour-branes will lead to the fundamental hypermultiplets.

Recall from Section \ref{ftheorysec} that in the IIB picture the D7--branes
were located on the orientifold plane defined by $(x^8,x^9)\ra -(x^8,x^9)$.
We will similarly take the flavour-branes to lie on the fixed point set of
our orientifold action $(\psi^A\ra-\psi^A)$, by imposing Dirichlet conditions
in the $\psi^3,\psi^4$ directions. We will also keep the Dirichlet condition on
the antiholomorphic $\bar{\psi}^{\bar{A}}$ directions. Since these new branes still
extend along the bosonic directions
of $\Supertwistor$ (as well as the fermionic $\psi^a$ directions), from now on
we will drop the possibly misleading `D5' terminology and label the branes
discussed in the last section (which led to the gauge group $\Sp(N)$) as `$\Dc$'
(for colour) and the new branes as `$\Df$' (for flavour). We summarise the boundary
conditions satisfied by open strings stretching between the branes in our setup
in Table \ref{Bcs}.

\begin{table}[h]
\begin{center}
\begin{tabular}{|c|c|c|c|} \hline
Direction & $\Dc$--$\Dc$ & $\Dc$--$\Df$ & $\Df$--$\Df$ \\ \hline
$Z$,$\Zbar$ & NN & NN & NN  \\
$\psi^a$ & NN & NN & NN  \\
$\psi^A$ & NN & ND & DD \\
$\bar{\psi}^{\bar{a}}$,$\bar{\psi}^{\bar{A}}$ & DD & DD& DD \\ \hline
\end{tabular}
\end{center}
\caption{Boundary conditions for open strings in the B--model
  setup.} \label{Bcs}
\end{table}

Having chosen the boundary conditions defining a $\Df$ brane, we will now need
to decide on a) how many of them to introduce and b) how the orientifold and orbifold
groups act on the Chan--Paton indices associated with these branes. For the
first question, it turns out that (as will become clear shortly) introducing
two $\Df$ branes, which along with their mirrors lead to a $4\times 4$ Chan--Paton group,
is what is necessary to reproduce the $N_f=4$ theory. We will call the corresponding
indices $X,Y,\ldots=1,\ldots,4$. As for the second question, recall that for the
$\Dc$ branes we chose the orientifold action $\tilde\gamma_c=i\Omega_{2N\times2N}$, but
the action of the orbifold was trivial: $\gamma_c=\mathbb{I}_{2N\times2N}$. With an
eye to the results we want to obtain, we will again choose the orientifold action antisymmetric
($\tilde\gamma_f=i\Omega_{4\times4}$), but this time we take $\gamma_f=-\mathbb{I}_{4\times4}$.
Thus, the full specification of
our orientifold action (extending (\ref{superorientifold})) is given by:
\be
\label{fullsuperorientifold}
\begin{split}
(a)&\quad  \psi^a\ra \psi^a \quad,\quad \psi^A\ra -\psi^A \\
(b)&\quad \mathcal{J}\ra \Omega_{(r)}\Jcal^{(T)}\Omega_{(r)} \\
(c)&\quad \mathcal{J}\ra \gamma_{(r)}\mathcal{J}\gamma^{(-1)}_{(r)}\;,
\end{split}
\ee
where the generic B--model state $\Jcal$ can take any of the four possible choices of
Chan--Paton indices (\emph{i.e.}
$\Jcal^i_{\;\;j},\Jcal^i_{\;\;X},\Jcal^X_{\;\;\;\;i},\Jcal^X_{\;\;\;Y}$), $\gamma_{(r)}$
is either $\gamma_c$ or $\gamma_f$ depending on the index it is acting upon, and
similarly $\Omega_{(r)}$ corresponds to either $\Sp(N)$ or $\Sp(2)$.

This completes the definition of our proposal for the twistor dual of the $N_f=4$ theory.
Now let us check whether we can recover the expected spectrum on the spacetime side.
Of course the discussion in Section \ref{orientsec} remains unchanged, so we already know
that the $c-c$ strings of our construction reproduce the correct vector and
antisymmetric hypermultiplet spectrum.

First, we will look at the $f-f$ strings, which will provide us with information
on the Chan--Paton group corresponding to the four $\Df$ branes. We thus need
to confront the problem of interpreting the Dirichlet boundary conditions in
the holomorphic $\psi^A$ directions.
Unlike what happens for the antiholomorphic fermions, simply interpreting these as
imposing $\psi^A=0$ (so that observables do not depend on $\psi^A$) does not seem to
provide the correct degrees
of freedom. The resolution comes through realising that one has to apply a fermionic
analogue of dimensional reduction, which is part of a more general question of
properly defining sub-supermanifolds of supermanifolds. Some aspects of this,
which turn out to be sufficient for our purposes, have been discussed in
\cite{Saemann04}, whose approach we will follow (and where further references can
be found). In brief, the results of \cite{Saemann04} indicate that a reasonable
definition of fermionic dimensional reduction is to restrict the fermionic dependence
of the original supermanifold so that fields on the sub-supermanifold can only
depend on them in certain combinations. For example, one of the cases considered
in \cite{Saemann04} was the reduction $\Supertwistor\ra \Cset\mathrm{P}^{3\oplus1|0}$,
where the notation \cite{KonechnySchwarz97} means that all four $\psi^I$
have been combined into a single
nilpotent bosonic coordinate $y = \psi^1\psi^2\psi^3\psi^4$.\footnote{$\Cset\mathrm{P}^{3\oplus1|0}$
is an example of a \emph{thickening} of $\Twistorspace$ \cite{Saemann04}.}

A simple way to impose such constraints on the fermionic dependence is in
terms of a suitable set of integral constraints, and indeed the particular
reduction above was first performed in \cite{LechtenfeldPopov04} using such
an approach. However, with this choice (as well as another case
considered in \cite{Saemann04}) one is
led to a completely bosonic truncation of the $\Ncal=4$ spectrum, while our
$\Df$ branes are still expected to preserve $\Ncal=2$ supersymmetry, so we will
need to slightly adapt those embeddings to our setting. Given the symmetries
of our system, we propose that the supermanifold reduction defining the $\Df$ branes is
$\Supertwistor\ra\Cset\mathrm{P}^{3\oplus1|2}$, where the nilpotent coordinate is
$\psi^3\psi^4$ and the $\psi^1$,$\psi^2$ coordinates are unrestricted.\footnote{Such
maps of supermanifolds, where one exchanges pairs of odd coordinates for even
nilpotent coordinates, have also appeared, in a slightly different (superspace) context,
in \cite{Harnadetal89}.}

As discussed above, the NN directions will provide a (0,1)--form
living on the $\Df$ branes, which we denote by $\Kcal^X_{\;\;Y}$.
The above definition of dimensional reduction can
be implemented by imposing the following eight equations (which are a subset of the
truncation conditions considered in \cite{LechtenfeldPopov04})
\be \label{truncation}
\int\diff^4\psi \psi^1\psi^2\psi^A\Kcal=
\int\diff^4\psi\psi^a\psi^A\Kcal=
\int \diff^4\psi \psi^A\Kcal=0\;.
\ee
These conditions restrict the $\psi$ dependence of $\Kcal$ to take the
following form
\be
\Kcal^X_{\;\;Y}=\diff\Zbar^{\bar{m}}\left({K(Z,\Zbar,\psi^a)_{\bar{m}}}^X_{\;\;Y}
+\psi^{3}\psi^4{L(Z,\Zbar,\psi^a)_{\bar{m}}}^X_{\;\;Y}\right)\;.
\ee
It is easy to check that requiring invariance under the orientifold action results in
a symmetric truncation of the Chan--Paton matrix defined
by the $X,Y$ indices and thus $\Kcal$ is a $4\times 4$ matrix
transforming in the adjoint of an $\Sp(2)$ group. Thus we have specified the (0,1)--form
part of the $f-f$ spectrum.

However, as can be seen in (\ref{VO}), the existence of holomorphic DD directions implies
that the $(0,1)$--forms do not exhaust the possible vertex operators that can be written
down at ghost number one. One can now also have states of the form
\be
\Bcal^A\theta_A \;\sim\; \Bcal^A(Z,\bar{Z},\psi^a,\psi^A)\frac{\partial}{\partial\psi^A}\ .
\ee
Motivated by dimensional reduction in the physical string case, and in particular
by the desire to have the same counting of states before and after the reduction,
we will assume that the fermionic dependence of these DD $f-f$ states arises by
considering the \emph{complement} of the eight equations in
(\ref{truncation}).\footnote{This becomes clearer if one chooses to reduce along
all four $\psi^I$ directions, as in \cite{LechtenfeldPopov04}. In that case one
imposes 14 equations in the NN sector, so the (0,1)--strings provide just
two degrees of freedom. The remaining states should then arise from the DD sector,
therefore we would want to impose just two equations on that sector.}
This will restrict the general expansion for $\Bcal$ to
\begin{eqnarray}
 \Bcal^A{(Z,\bar{Z},\psi)^X}_Y\frac{\partial}{\partial\psi^A}=
\psi^B B_B^A{(Z,\bar{Z},\psi^a)^X}_Y\frac{\partial}{\partial\psi^A}\ .
\end{eqnarray}
Requiring invariance under the orientifold action (under which we also have
$\partial/\partial\psi^A \ra -\partial/\partial\psi^A$) once again restricts the
Chan--Paton indices to be those of $\Sp(2)$. It is straightforward to check that
$\psi^B B_B(Z,\Zbar,\psi^a)$ provides 4 fermionic and 4 bosonic degrees of freedom,
which, together with $\Kcal$, give the expected counting of states for the 8d $\Ncal=1$
theory on the D7--brane (note that in this counting we suppress the index corresponding
to the expansion of $B$ in a basis of $T^{(1,0)}$, in the same way that we have been
suppressing the form index $\bar{z}$ for the (0,1)--form states). These states,
not being (0,1)--forms, are clearly unsuitable for a straightforward application of the
Penrose transform to four dimensions. This is not unexpected, since their natural
dual interpretation would be as states of the eight--dimensional D7--brane theory.
We will further comment on such a potential interpretation at the end of this section.

It should also be pointed out that, again because they are not (0,1)--forms,
there seems to be no obvious way to include the $\Bcal$ states in a
holomorphic Chern--Simons--type action (which would still need to be integrated
over a (3,3)--cycle), and in particular we cannot write
down the action on the $\Df$ worldvolume including these terms
by dimensional reduction (unlike the case for bosonic DD directions, see \eg
\cite{Hofman02}). Perhaps a suitable generalisation of the hCS action, along with
a more rigorous definition of our integration measure, would be able
to accommodate this more general case, but since for the purposes of this paper
we will only need to know the $\Dc$ brane action, which is what is expected
to have a relation to the 4d theory that we are interested in, we will not pursue
this question further here.

 Clearly the choice of the above geometric embedding of the $\Df$ branes
within $\Supertwistor$ has been based on rather heuristic arguments, and,
although it certainly seems to provide a consistent picture, we cannot claim
that it is the unique possibility. It would certainly be desirable to obtain
a more fundamental understanding of this embedding starting from the basic
definition of Dirichlet boundary conditions on the B--model worldsheet. Leaving
this for future work, we will now turn to the last aspect of our construction,
\ie\: the strings stretching between the $\Dc$ and $\Df$ branes.

Therefore, we finally consider the $c-f$ and $f-c$ strings. Recall that these
are the real reason to introduce the $\Df$ branes, since they will provide the
desired fundamental matter. Looking at Table \ref{Bcs}, and recalling that
(topological) DN strings do not have zero modes and thus do not provide B--model
states, the only contributions arise from the NN sector.
Suppressing the $(0,1)$--form index, these can be usefully written as an
expansion in $\psi^A$
\be\label{bosonicQs}
\Qcal^i_{\;\;X}=P(Z,\Zbar,\psi^a)^i_{\;\;X}+\psi^AQ_A(Z,\Zbar,\psi^a)^i_{\;\;X}
+\psi^3\psi^4 R(Z,\Zbar,\psi^a)^i_{\;\;X}
\ee
and similarly for the $f-c$ field $\Qcal^X_{\;\;i}$. Note that, due to the
orientifold action (\ref{fullsuperorientifold}.b), the $c-f$ and $f-c$ states are
related by the condition
\be \label{condition}
\Qcal^X_{\;\;i}=\Omega_{ij}\Qcal^j_{\;\;Y}\Omega^{YX}\;.
\ee
It is easy to check that the other components of (\ref{fullsuperorientifold})
impose $P^i_{\;\;X}=R^i_{\;\;X}=0$ and thus dictate that the $c-f$ and $f-c$ states are given by
\be\label{Qcal}
\Qcal^i_{\;X}=\psi^A Q^i_{AX}\qquad , \qquad \Qcal^X_{\;\;\;i} = \psi^A Q_{A\;i}^{X} \; ,
\ee
where we can expand
\be\label{Q}
Q^i_{AX} = \eta^i_{AX}+ \psi^aq^i_{aAX}+ \psi^1 \psi^2
\tilde\eta^i_{AX} \;
\ee
and similarly for $Q^X_{Ai}$. Recall that here $i$ is an $\Sp(N)$
gauge group index,  $A$ is an index of $\SU(2)_A$  and (as we previously derived)
$X$ is an index of $\Sp(2)$. The particular form of $Q$ is not new: As shown
in \cite{Ferber78,Boelsetal06}, this is the precise twistor field content
(for each value of $X$)
corresponding to an $\Ncal=2$ hypermultiplet!\footnote{To be more precise, these
references describe a hypermultiplet as consisting of two fermionic half--hypermultiplets,
while in our case they naturally appear in $\SU(2)_A$ doublets, at the cost of losing
manifest $\SO(8)$ invariance.}  We conclude (and will make more
precise shortly) that our orientifolding
procedure has produced a hypermultiplet $Q^i_{AX}$ in the fundamental representation of
$\Sp(N)$.

Let us now investigate its transformation properties under the two
global groups, given by the indices $A$ and $X$. As we reviewed
in Section \ref{spacetime}, the
fundamental hypermultiplets should also transform in the fundamental representation of
the global $\SO(8)$ flavour group. However at the end of that section we explicitly
decomposed the $\SO(8)$ into its $\SU(2)\times\Sp(2)$ subgroup.
The reason for that should now be evident: In the twistor string model we have
constructed, the SU(2) arises geometrically as the symmetry under which the
$\psi^A$ coordinates transform as doublets, while the remaining $\Sp(2)$ arises
as the Chan--Paton group of the flavour--branes.
We will explore some of the implications of this decomposition of the flavour
group shortly, but it is clearly an unavoidable consequence of the fundamental fields
in (\ref{Qcal}) being linear in $\psi^A$. However, we
can immediately comment on another consequence of this linear behaviour:
It provides a very natural explanation for the fermionic nature of
$Q^i_{AX}$ which had to be assumed in the constructions of \cite{Ferber78,Boelsetal06}.

We conclude that, by defining our flavour-branes to lie at the orientifold
fixed point, and extending the orientifold action to act nontrivially
on their Chan--Paton indices, we have reproduced the fundamental part of
the spectrum of the $N_f=4$ theory. This description has several peculiarities
relative to the physical string description, not least of which is the
fact that the relative sizes of the D3 and D7 branes in the IIB setup seem to
be interchanged: Our $\Df$ branes extend (have NN b.c.'s) along a
subspace of that of the $\Dc$ branes and could perhaps be thought of as
defects in the worldvolume theory of the latter. On the other hand, what is
perhaps more relevant in comparing to the spacetime picture is the \emph{super-dimension}
of our branes, defined as the difference between the number of bosonic and fermionic NN
directions.\footnote{For instance, \cite{Schwarz95} argues for the equivalence
of the A--model on certain $(m|n)$--dimensional supermanifolds to that on bosonic
$(m-n)$--dimensional manifolds. See also \cite{Ricci05} for similar observations in the
context of mirror symmetry.} Although this deserves
further study, we note that it also seems to be consistent with an observation
in \cite{Tokunaga05} that (for non--topological strings on supermanifolds)
the number of  fermionic NN directions contributes to the brane tension
inversely to that of bosonic NN directions, and thus a brane extending along
fewer fermionic directions can be thought of as having larger mass. Although
these results do not apply directly in our setting, we take them as an
indication that the geometric embedding of the $\Df$ branes is the correct one.

Another perhaps surprising feature of our model is the fact that both
the $\Dc$ and $\Df$ branes were chosen to satisfy symplectic projection
conditions on their Chan--Paton indices, leading to $\Sp(N)$ and $\Sp(2)$
worldvolume gauge groups respectively. This seems to conflict with the arguments
of \cite{GimonPolchinski96} which (applied to the orientifolded D3--D7 system)
would require opposite projections for the two types of branes, leading to
$\Sp(N)$ and $\SO(8)$ gauge groups. However, that analysis was based on subtle
properties of the $3-7$ string DN sector, which is absent in this
case. Therefore it would seem that the B--model is too simple to accommodate
such an effect, but confirmation of this will have to wait for a better
worldsheet understanding of our orientifold prescription.\footnote{
It is likely that the notions of B--parity and B--orientifolds, developed
for (untwisted) (2,2) models in \cite{BrunnerHori03} (see also \cite{HoriWalcher06}),
properly extended to the supermanifold case, will be of help in this regard.}

 Given that, in the physical string setup, our $\Df$ branes correspond to IIB
D7--branes, with an associated eight--dimensional worldvolume SYM theory,
it is fascinating to speculate that our twistor string model might, via a
suitable higher--dimensional generalisation of the Penrose transform, also have
another dual description in terms of an \emph{eight--dimensional} spacetime theory.
Under this duality, the worldvolume theory of the twistor $\Df$ brane would presumably
map to some integrable subsector of 8d Yang--Mills. A preliminary remark in this
direction is that a natural definition of selfduality
for 8d Yang--Mills \cite{Ward84} also seems to require the same breaking of
(Lorentz) $\SO(8)$ to $\Sp(2)\times\Sp(1)$ that we observe on the twistor
side. Although it would be very interesting to understand this connection better,
we will from now on focus on the standard four--dimensional Penrose transform that
connects the spectrum and field equations of the $\Dc$ brane worldvolume
theory to those of a suitable generalisation of 4d selfdual
Yang--Mills.\footnote{In doing this we will assume that the Penrose transform
can be applied just to the $\Dc$ brane theory, comprising the $c-c$ strings plus
their interactions with the $c-f$ and $f-c$ strings, ignoring interactions
 with the $\Df$ worldvolume theory. In the physical string
setting such interactions are frozen at low energies essentially due to the
difference in spatial extent of the D3 and D7--branes. It would be interesting
to identify a mechanism providing such a decoupling in our topological string
setting.}

\subsection{The Final Twistor Action}

In the last two sections we defined a B--model setup with certain numbers of
branes that reproduced the spectrum of the $N_f=4$ theory. The resulting
superfields can be naturally embedded into the holomorphic Chern-Simons
action in the following way\footnote{Here we write the fundamental part of
the action by analogy with that for the antisymmetric fields. However, note the
different relative coefficient of the interaction terms, which is due to their
different $\Sp(N)$ transformation properties.}
\begin{eqnarray} \label{Nf=4TotalAction}
S &=& \frac{1}{2}\int_{\Dc}\mathbf{\Omega} \wedge
\left(\mathrm{Tr}[\hat {\mathcal A}\cdot\bar\partial \hat{ \mathcal A}+
  \frac{2}{3} \hat{ \mathcal
A}\wedge\hat{\mathcal A}\wedge \hat{ \mathcal A}]+ \mathcal Q^{ X}\cdot
\bar \partial
\mathcal Q_{ X}+\mathcal Q^{ X}\wedge \hat{ \mathcal
A} \wedge \mathcal Q_{ X} \right)\nn\\
&=& \frac{1}{2}\int_{\Dc}\mathbf{\Omega} \wedge\Bigg(\mathrm{Tr}[\Vcal\cdot\dbar\Vcal+
\frac{2}{3}\Vcal\wedge\Vcal\wedge\Vcal+\Zcal\cdot\dbar\Zcal+2\Zcal\wedge\Vcal\wedge\Zcal]\nn\\
&&\qquad\qquad\qquad+\; \mathcal Q^{ X}\cdot
\bar \partial
\mathcal Q_{ X}+\mathcal Q^{ X}\wedge \Vcal \wedge \mathcal Q_{ X}\Bigg)\ .
\end{eqnarray}
The classical equations of motion can then be easily found to be
\bea
\nn \dbar \Vcal +\Vcal\wedge\Vcal+\Zcal\wedge\Zcal+\frac{1}{2}\Qcal^X\wedge\Qcal_X &=& 0\\
\nn \dbar \Zcal +[\Vcal,\Zcal]&=&0\\\
\dbar \Qcal_X +\Vcal\wedge\Qcal_X &=& 0
\eea
and by linearising these around the trivial solutions $\Vcal = 0$, $\Zcal=0$, $\Qcal = 0$
one obtains
\begin{eqnarray}
\label{closed}
 \dbar \Vcal = \dbar\Zcal = \dbar \Qcal = 0\ .
 \end{eqnarray}
In addition, (\ref{Nf=4TotalAction}) has the following three gauge invariances, related
to three different $(0,0)$-form gauge parameters $\gre^i_{\;\;j}$, $\varepsilon^i_{\;\;j}$
and $e^i_{\;\;X}$
\be
(a)\;\delta\Vcal=\dbar\gre+[\Vcal,\gre]\;,\quad
\delta\Zcal=[\Zcal,\gre]\;,\quad
\delta \Qcal^X_{\;\;i}=\Qcal^X_{\;\;j}\gre^j_{\;\;i}\;,\quad
\delta \Qcal^i_{\;\;X}=-\gre^i_{\;\;j}\Qcal^j_{\;\;X}\;,
\ee

\be
(b)\;\delta\Zcal=\dbar \varepsilon+[\Vcal,\varepsilon]\;,\quad
\delta\Vcal=[\Zcal,\varepsilon]\;,
\ee
and
\be
(c)\;\delta \Qcal^i_{\;\;X}=\dbar e^i_{\;\;X}+\Vcal^i_{\;\;j} e^j_{\;\;X}\;,\quad
\delta \Qcal^X_{\;\;i}=\dbar e^X_{\;\;i}-e^X_{\;\;j}\Vcal^j_{\;\;i}\;,\quad
\delta\Vcal^i_{\;\;j}=\half(\Qcal^i_{\;\;X} e^X_{\;\;j}-e^i_{\;\;X}\Qcal^X_{\;\;j})\;.
\ee
The first of these is the ordinary gauge invariance while the other two are
clearly very unusual, and are due to the fact that on the twistor side $\Zcal$ and
$\Qcal$ are (0,1) forms.\footnote{In fact (a) and (b) can be straightforwardly derived
from the transformation of $\hat\Acal$ ($\delta\hat\Acal=\dbar E+[\hat\Acal ,E]$),
by splitting $\hat\Acal=\Vcal+\Zcal$ and $E=\gre+\varepsilon$ into symmetric and
antisymmetric parts and considering the symmetry properties of the resulting terms.}
Essentially the same transformations have been discussed in \cite{Boelsetal06},
where they arise as symmetries of the (non-cubic) twistor space effective action
which, in the formalism there, would correspond to full (non-selfdual) $\Ncal=2$ SYM
with matter.

As such, the linearised equations of motion and these symmetries are enough to put the
superfields $\Vcal,\Zcal$ and $\Qcal$ in the appropriate cohomology classes for their
component fields to map to spacetime states. In particular,
the components of $\Qcal$ then map to Minkowski space fields of helicity $(\frac{1}{2},0,-\frac{1}{2})$ via the
Penrose transform
\be\label{Qspectrum}
\nn\mathcal Q \quad=\underbrace{(0,\eta_{AX},q_{aAX},\tilde{\eta}^{AX} ,0)}_{\textrm{\small{1--forms of S-charge $(-k)$ in twistor space}}}\stackrel{\textrm{\small{Penrose}}}\longleftrightarrow\underbrace{(0,\eta_{AX},q_{aAX},\bar{\eta}^{AX} ,0)}_{\textrm{\small{fields of helicity $(1-k/2)$ in Mink. space}}} \;.
\ee

We have thus obtained the complete spectrum of the $N_f =4$ theory from twistor string theory.
Expanding (\ref{Nf=4TotalAction}) in  components and integrating out the
fermionic variables gives
\be\begin{split}
  S_{hCS}= & \int_{\mathbb{C}\mathrm{P}^3}\mathbf{\Omega'}\wedge \left(\mathrm{Tr}[G\wedge {F}+
\phi^{\dag}\wedge\bar{D}\phi
-\tilde{\lambda}^a\wedge\bar{D}\lambda_a
+\lambda^a\wedge\lambda_a\wedge\phi^{\dag}\right. ]\\ & +\Tr[-\frac{1}{2}z^{aA}\wedge\bar{D}z_{aA}
-\tilde{\zeta}^A\wedge\bar{D}\zeta_A
-z^{aA}\wedge\lambda_a\wedge\zeta_A+\zeta^A\wedge\zeta_A\wedge\phi]\\ &\left. + \tilde \eta_{AX}\wedge\bar
 D\eta^{AX}
-\frac{1}{2}q_{aAX}\wedge\bar D q^{aAX}- q_{a A
 X}\wedge \lambda^a \wedge \eta^{AX} +\frac{1}{2}
 \eta_{AX}\wedge \phi\wedge \eta^{AX}\right)\; ,
\end{split}
\ee
where the covariant derivatives are defined as $\bar D  =
\bar\partial + [A,\;]$ for tensor fields and $\bar D=\dbar+A\wedge$ for fundamental ones.
This looks very much like the selfdual truncation of the $N_f = 4 $ theory that
we obtained in (\ref{selfdualsp}), which we present again  to facilitate the comparison
\be
\nn \begin{split}
S_{\textrm{4}d} =\int d^4x \; & \mathrm{Tr} \left[ -\frac{1}{2}  GF  +
   D\phi^{\dagger} D\phi + i \bar \lambda^a \Dslash \lambda_a
 -    \lambda^a\lambda_a \phi^{\dagger}   \right] \\
&     - \mathrm{Tr} \left[\frac{1}{2}
   D z^{aA}D z_{aA}  + i \bar\zeta^A  \Dslash
\zeta_A  +  z^{aA}[\lambda_a,
 \zeta_A] + \zeta^A\zeta_A \phi \right]\\
&  -\left( \frac{1}{2}  Dq_{a A' X} Dq^{ a A' X} + i \bar \eta_{A' X}\Dslash
\eta^{A' X}   +
   q_{a A' X} \lambda^a \eta^{A' X}  + \frac{1}{2}\eta_{A' X}\phi\eta^{A' X}\right)
 \; . \end{split}
\ee
As we have already mentioned, there should exist a nonlinear
generalisation of the Penrose transform
in the spirit of \cite{PopovSamann04}, relating these two actions exactly.
Moreover, note that by comparing the two we readily observe that even
though there is both an $\SU(2)_A$ and an $\SU(2)_{A'}$ symmetry for the gauge
theory, we only see a single $\SU(2)_A$ on the B--model side. This is a hint that these
two symmetries are identified in the twistor string description,
a claim which we will verify during the  comparison of amplitudes between
the two theories.

In summary, we have introduced four $\Df$ branes parallel to the
superorientifold plane which account for the $\Sp(2)$ part of
the flavour symmetry. Via
the Penrose transform, this yields the right spectrum for the
fundamental hypermultiplets in the $N_f = 4$ theory and mimics the
behaviour of the D7--branes in the physical string setup.
As we further discuss in the conclusions, it
would be intriguing if there were a mechanism which exactly fixes the
number of $\Df$ branes in the B--model to four (two plus two mirrors),
\eg some analogue of the RR charge cancellation condition in string theory.
The existence of such a mechanism would suggest (as expected perhaps) that
our construction is only consistent
at loop level for the precise case when the dual gauge theory is finite.

\section{Comparison of amplitudes} \label{Amplitudes}

Having reproduced the spectrum of the $N_f=4$ theory, we will now establish the duality on firmer grounds by calculating amplitudes in both the
gauge  theory and  topological string theory, and by showing precise agreement (up to a constant
normalisation factor).

\subsection{Review of the standard amplitude prescription} \label{N4review}

We will begin by briefly summarising the prescription of \cite{Witten0312}
for the calculation of colour-stripped partial amplitudes in $\Ncal=4$ SYM.
As we indicated above, this reduces to the evaluation of particular correlators on
the worldvolume of D1--instantons wrapping curves of a certain degree in
$\Supertwistor$ and then integrating over the moduli space of such curves.
For tree--level MHV amplitudes, the D1--instantons are localised \cite{Witten0312}
on $\Cset\mathrm P^1$s in $\Supertwistor$ with the embedding given by
\be \label{embedding}
\mu_{\dot \alpha} + x_{\alpha\dot \alpha} \lambda^\alpha = 0 \quad
\textrm{and} \quad \psi^I +\theta_\alpha^I \lambda^\alpha = 0\;,
\ee
where $Z^m = (\lambda^\alpha , \mu^{\dot \alpha})$ and $\psi^I$ are the
supertwistor space coordinates, while the moduli
$x_{\alpha\dot \alpha}$ and $\theta^I_\alpha$ correspond to the coordinates of
4d Minkowski space and (on--shell) $\Ncal=4$ superspace respectively.

Following an idea due to Nair \cite{Nair88}, the gauge theory amplitudes
are reproduced by correlation functions of chiral currents on the
worldvolume of these  D1--instantons. Since the insertion of
these objects explicitly breaks the isometries of $\Supertwistor$,
one must integrate over the moduli space of instantons of the  appropriate
degree. The prescription for the
calculation of tree-level MHV amplitudes, and therefore integration over degree
one, genus zero curves,
is then
\be\label{amp}
A_{(n)} = g^2 \int d^4x\; d^8\theta \;\langle \int_{\MHV}\!\!
J_1w_1\cdots \int_{\MHV}\!\! J_nw_n \rangle\;,
\ee
where $J_i$ are D1 worldvolume free--fermion currents coupling to
the external D5--brane fields (including both the colour and flavour-branes
in our case), while the $w_i$'s are the
twistor space equivalents of wavefunctions for the external particles.
The lower index $i=1,\dots ,n$ indicates the position of the external particle
in the $n$-point scattering process, as well as the point onto
which these localise on the holomorphic curve in twistor space.
The factor of $g^2$ is identified with the D1-instanton
expansion parameter.
The calculation for the product of the currents boils down to yielding a
gauge group factor, which we will strip off,  as well as the following
denominator part of the MHV amplitudes\footnote{Here we use the widespread
notation $\langle 1 2 \rangle = \langle \lambda_1 \lambda_2 \rangle
=  \lambda_1^\alpha \lambda_{2\alpha}$ and $[12] =
[\tilde\lambda_1\tilde\lambda_2] =   -\tilde\lambda_{1\dot
  \alpha}\tilde\lambda_2^{\dot \alpha}$, with $2 (p_i\cdot p_j) = \langle \lambda_i \lambda_j \rangle[\tilde\lambda_i\tilde\lambda_j]$.
  See also Appendix \ref{Feynman}.}
\be
\langle J_1\cdots J_n \rangle_{\textrm{\small{stripped}}} = \frac{1}{\langle 12 \rangle \langle 23 \rangle\ldots\langle n 1 \rangle} \;.
\ee
The numerator of the amplitude is produced by the twistor wavefunctions $w_i$,
which, upon integration over the positions of vertex operators for
each on-shell external particle, result in a colour-stripped coefficient $v_i(\psi_i)$ equal to the
one
in the superfield expansion of $\Acal$ in (\ref{N=4 superfield}) \cite{CachazoSvrcek05}.
These contribute a number of factors of $\psi$, which are then integrated over the
moduli space of D1--instantons via the embedding relation $\psi_i^I = \theta^{\alpha I}\lambda_{i\alpha}$.
Since the fermionic part of the measure on moduli space for genus zero, degree one holomorphic curves is
$d^8\theta$, the MHV amplitude is non-zero only if the Grassmann integral is saturated, that is,
 if the total S-charge of the external states participating in the
 scattering process is $\mathrm S = -8$.
 Conversely, if a process involves external states with total charge $\textrm S = -8$,
 it is then MHV. Since in the case under study these amplitudes can include external fermions
 or scalars satisfying this condition in addition to gluons,
it is perhaps more appropriate to refer to them as `analytic' \cite{Khoze04} rather
than MHV, and we will mostly use the latter notation in the following.
Finally  we note that the integral over the bosonic moduli yields a $\delta$--function of
 momentum conservation, which we omit. This prescription successfully reproduces all amplitudes
localising on holomorphic, degree one, genus zero curves in $\Ncal=4$ SYM.

The above can also be extended to amplitudes which localise on higher degree, genus
zero holomorphic curves. For generic  scattering states this degree is given by $d=-\frac{1}{4}\sum_{i=1}^n \mathrm S_i-1$, where the sum is over the
S-symmetry charges of the $n$ external particles. For gluon scattering these correspond to
next-to-MHV (NMHV) and higher $\mathrm{(N^{q-2}MHV)}$ amplitudes and
the appropriate degree is given by
$d = q-1$, where $q $ is the number of negative helicity
gluons. Although the original string--motivated prescription of \cite{Witten0312}
made use of one connected degree--$d$ instanton, in practice it turned out to be
more useful to consider instead a sum of $d$ disconnected (degree one) D1--instantons, leading
to the MHV--rules prescription \cite{Cachazoetal0403}. The equivalence of these
prescriptions (as well as intermediate pictures of multiple D1--instantons of degrees
adding up to $d$) is strongly suggested by the work of \cite{Gukovetal0404}.

\subsection{Extension to the $N_f=4$ theory}

The above prescription can be straightforwardly extended to the twistor model
for the $N_f=4$ theory that we constructed in Section \ref{Twistorsec}.
The starting point is to consider D1--instantons localised along holomorphic
curves in the orientifold of $\Supertwistor$, which now includes the two types of
`D5' branes, which we have denoted $\Dc$ and $\Df$.  Assuming that the
D1 worldvolume currents couple to the external $\Df$ fields in the same way as
to the $\Dc$'s, we will take the formula (\ref{amp}) as our starting point.
 The difference in this case is that the twistor wavefunctions $w_i$
will now associate the appropriate term
in the superfield expansion of the
$\mathcal V , \mathcal Z$ of (\ref{VZ}) and
$\Qcal$ of (\ref{Q}) with each on-shell external particle.\footnote{The reader  worried about only
integrating over the moduli space of $\Cset \mathrm{P}^1$s in an
orientifolded theory, which should  also include
$\mathbb{R}\textrm{P}^2$ topologies \cite{Cicuta82}, should recall that these
contributions are non-planar and will be absent at tree-level. They
should, however, play a role in any eventual loop level calculation.}

The fact that the gauge group is now $\Sp(N)$ rather that $\SU(N)$ does not
introduce major complications, due to the fact that we consider colour stripped
partial amplitudes, effectively
factoring out all information about the gauge group. In the usual approach to organising
amplitudes in  $\U(N)$ gauge theories,\footnote{See for example the reviews
\cite{ManganoParke91,Dixon96}.} this amounts to considering
definite orderings for the external scattering states and then summing over
all non-cyclic permutations to obtain the full amplitude. The structure
of the group theory piece leads to identities, which dramatically simplify
the calculation by allowing the evaluation of a great number of
partial amplitudes  by simply exchanging negative
helicity spinor factors. A similar procedure can be
applied to the $\Sp(N)$ case. Naturally, from a given colour stripped
result, one can recover  different full amplitudes depending on the
gauge group choice. Since $\Sp(N)$ gauge theory amplitudes seem to  have no real
phenomenological importance and since agreement of partial amplitudes
between the gauge and twistor theory sides is enough to
establish their correspondence, we will not explicitly calculate the
full answer, although it is straightforward to recover it  using
simple  group theory facts.\footnote{Pseudoreality of $\Sp(N)$ will
  make this step slightly more subtle compared to $\U(N)$, since
  there exist extra identities relating different orderings of the
  external particles.}
We would like to note at this point that we will not only strip the gauge group
indices but also the $\Sp(2)$ indices $X$, which appear in the
definitions of the
fundamental fields. The motivation for this is that they are global non-geometric
indices and the
partial amplitude calculation is insensitive to how one chooses to contract them. In obtaining the full amplitude involving external fundamental fields, one should of course be careful to properly consider all possible contractions that lead to an $\Sp(2)$ scalar quantity.

In order to demonstrate that the standard twistor prescription for tree--level
analytic amplitudes can be applied, essentially unmodified, to the $N_f=4$ theory,
we will now move on to explicit calculations of partial
amplitudes. We will do this for a large set of amplitudes of different combinations
involving  external particles transforming in the adjoint,
antisymmetric and fundamental representations of the $\Sp(N)$ gauge
group.\footnote{We do not need to calculate gluon scattering processes since the
stripping procedure guarantees that the partial amplitudes will go through as in
the $\Urm(N)$ case.}  The first nontrivial analytic amplitudes appear at 4-point but
we will also evaluate a few 5-point amplitudes to provide further
evidence for the duality. In the following subsections we will
explicitly display  the result on the twistor string side. In order to
get the result purely from gauge theory one needs to extract
the Feynman rules from the Lagrangean (\ref{Nf4full})
and then add up the contributions
from all channels for the process under consideration.
In Appendix \ref{Feynman} we list these Feynman rules in spinor helicity formalism,
as well as various identities we have employed in order to obtain the spacetime
answer. Since we do not have a precise map between the actions on the two
sides of the correspondence, we cannot hope to exactly match the
resulting amplitudes. We therefore calculate ratios of the latter  and
find exact agreement up to a relative constant normalisation
factor. In particular, in our conventions we find that the spacetime answer is
obtained from the twistor result by
multiplying by a factor of $32i$.

\subsection{`Pre-analytic' amplitudes}
Before proceeding with the analytic results, we will briefly look at
the amplitudes that have a total value of $\mathrm S = -4$, which we will
call pre-analytic. These are $\langle \lambda^a , \lambda^b , \eta_A ,
\eta_B\rangle$, $\langle \lambda^a ,  \eta_A , \lambda^b ,
\eta_B\rangle$, $\langle \lambda^a , \lambda^b , \zeta_A ,
\zeta_B\rangle$ and  $\langle \lambda^a ,  \zeta_A , \lambda^b ,
\zeta_B\rangle$ and on the
twistor side they correspond to amplitudes that localise on degree
zero curves in twistor space, \emph{i.e.} points. This means that all
particles are attached to the same point in twistor space and
$\lambda_i = \lambda_j$ $\forall\, i,j$. Therefore $2(p_i\cdot p_j)
= \langle \lambda_i \lambda_j \rangle[\tilde\lambda_i\tilde\lambda_j]
= 0$, and thus scattering amplitudes with $n\ge 4$, which depend on
such nontrivial kinematic invariants, must vanish \cite{Witten0312}.

 From the spacetime point of view this result is less obvious and one needs
to calculate all the corresponding amplitudes explicitly.
These come from interaction vertices which originate exclusively
from the selfdual truncation of the $N_f = 4$ theory (\ref{selfdualsp}).
In fact, this observation extends to all other theories admitting a tree-level twistor
string description. Moreover, since we only focus on the colour-stripped
(and $\Sp(2)$-stripped) partial amplitudes, it suffices to calculate processes
involving either fundamental or antisymmetric matter fields;
the amplitude is insensitive to their gauge transformation properties.
We will therefore only discuss the following examples involving the
fundamental fermions $\eta$.
\begin{description}
\item[A.]{\bf The amplitude $\langle \lambda_1^a , \lambda_2^b , \eta_{A,3} , \eta_{B,4}\rangle$}

There are two channels contributing to this amplitude, namely
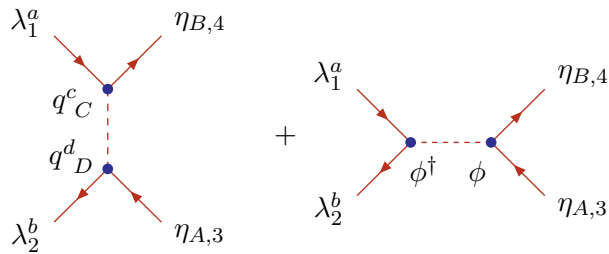
\be
\nn \begin{picture}(100,60)(-10,40)
\put(0,0){
\SetColor{BrickRed}
\ArrowLine(10,80)(30,60)
\ArrowLine(30,60)(50,80)
\DashLine(30,60)(30,30){3}
\ArrowLine(30,30)(10,10)
\ArrowLine(50,10)(30,30)
\SetColor{Blue}
\Vertex(30,60){2}\Vertex(30,30){2}
\Text(5,85)[r]{$\lambda^a_1$}
\Text(55,85)[l]{$\eta_{B,4}$}
\Text(25,55)[r]{$q^c_{\phantom c C}$}
\Text(25,35)[r]{$q^d_{\phantom d D}$}
\Text(5,5)[r]{$\lambda^b_2$}
\Text(55,5)[l]{$\eta_{A,3}$}
}
\end{picture}
+
\begin{picture}(120,60)(-10,30)
\put(0,0){
\SetColor{BrickRed}
\ArrowLine(10,50)(30,30)
\ArrowLine(60,30)(80,50)
\DashLine(30,30)(60,30){2}
\ArrowLine(30,30)(10,10)
\ArrowLine(80,10)(60,30)
\SetColor{Blue}
\Vertex(30,30){2}\Vertex(60,30){2}
\Text(5,55)[r]{$\lambda^a_1$}
\Text(85,55)[l]{$\eta_{B,4}$}
\Text(35,25)[t]{$\phi^\dagger$}
\Text(55,23)[t]{$\phi$}
\Text(5,5)[r]{$\lambda_2^b$}
\Text(85,5)[l]{$\eta_{A,3}$}
}
\end{picture}
\ee
\vspace{1cm}

One can easily verify by explicit calculation, using the Feynman rules provided in Appendix \ref{Feynman},
that they indeed cancel each other to give zero.

\item[B.]{\bf The amplitude $\langle \lambda_1^a ,  \eta_{A,2} , \lambda_3^b , \eta_{B,4}\rangle$}

The contributions to this process are

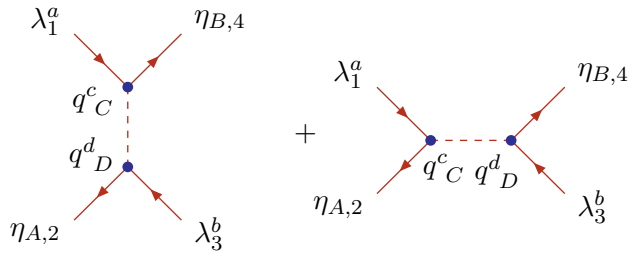
\be
\nn \begin{picture}(100,60)(-10,40)
\put(0,0){
\SetColor{BrickRed}
\ArrowLine(10,80)(30,60)
\ArrowLine(30,60)(50,80)
\DashLine(30,60)(30,30){3}
\ArrowLine(30,30)(10,10)
\ArrowLine(50,10)(30,30)
\SetColor{Blue}
\Vertex(30,60){2}\Vertex(30,30){2}
\Text(5,85)[r]{$\lambda^a_1$}
\Text(55,85)[l]{$\eta_{B,4}$}
\Text(25,55)[r]{$q^c_{\phantom c C}$}
\Text(25,35)[r]{$q^d_{\phantom d D}$}
\Text(5,5)[r]{$\eta_{A,2}$}
\Text(55,5)[l]{$\lambda^b_{3}$}
}
\end{picture}
+
\begin{picture}(120,60)(-10,30)
\put(0,0){
\SetColor{BrickRed}
\ArrowLine(10,50)(30,30)
\ArrowLine(60,30)(80,50)
\DashLine(30,30)(60,30){3}
\ArrowLine(30,30)(10,10)
\ArrowLine(80,10)(60,30)
\SetColor{Blue}
\Vertex(30,30){2}\Vertex(60,30){2}
\Text(5,55)[r]{$\lambda^a_1$}
\Text(85,55)[l]{$\eta_{B,4}$}
\Text(35,25)[t]{$q^c_{\phantom c C}$}
\Text(55,25)[t]{$q^d_{\phantom d D}$}
\Text(5,5)[r]{$\eta_{A,2}$}
\Text(85,5)[l]{$\lambda^b_{3}$}
}
\end{picture}
\ee
\vspace{1cm}

and similarly we find that after summing both parts the total vanishes.
\end{description}

This demonstrates (at four--point level) that all pre-analytic
amplitudes, which are the ones that can be constructed from the interactions in the
selfdual truncation of the theory, vanish after summation over channels.
The same phenomenon occurs for the selfdual truncation of $\Ncal = 4$ SYM \cite{Witten0312}.
In that case, as for the selfdual truncation of pure (non--supersymmetric)
Yang--Mills \cite{Yang77}, this fact is explained by noting that the
theory is classically integrable and is thus equipped with an infinite set
of (nonlocal) conserved charges.\footnote{For the pure selfdual YM case these can be found
(for instance) via
the Ward construction \cite{Ward77}; see \cite{Popov98} for a discussion and more
references. For $\Ncal=4$--extended selfdual YM an associated linear system was
discussed in \cite{Volovich83} and more recently its hidden symmetries were explored
in \cite{Wolf04}.} The corresponding Ward identities are then expected to constrain
tree--level amplitudes so severely that they are forced to vanish (brief discussions
on this can be found in \cite{OoguriVafa91b,Cangemi9605}). Thus, the vanishing of
pre--analytic amplitudes that we observe strongly suggests that
the selfdual sub-sector of the $N_f=4$ theory (which is a very different supersymmetric
extension of pure selfdual Yang--Mills from the $\Ncal=4$ case) also describes a
classically integrable system. It would be interesting to check this by explicitly
constructing the relevant conserved currents.

\subsection{The amplitudes
$\langle \phi, \phi, \phi^\dagger, \phi^\dagger\rangle$ and $\langle \phi, \phi^\dagger, \phi, \phi^\dagger\rangle$  }
We now turn to the analytic amplitudes of the theory. We start with
two simple examples involving only external adjoint scalars. There are
two possible orderings in this case and we will calculate both, to
show that these indeed give rise to different partial amplitudes. On
the  twistor side,  following the prescription (\ref{amp}) that we
have discussed in some detail, we can read off  and plug in the
wavefunctions appropriate to the $\langle \phi, \phi, \phi^\dagger,
\phi^\dagger\rangle$ amplitude from (\ref{VZ})
\bea
\nn v_1(\phi) = \psi^1_1\psi^2_1 &,& v_2(\phi) = \psi^1_2\psi^2_2\\
v_3 (\phi^\dagger)=\psi^3_3\psi^4_3 &,& v_4(\phi^\dagger) = \psi^3_4\psi^4_4\;.
\eea
The result is then given by the integral
\be
\langle \phi, \phi, \phi^\dagger, \phi^\dagger\rangle_{\mathrm{Twistor}}  = g^2\int d^8\theta \frac{\psi^1_ 1\psi^2_1\psi^1_2\psi^2_2\psi^3_3\psi^4_3\psi^3_4\psi^4_4}{\langle  12  \rangle \langle  23  \rangle\langle  34  \rangle\langle  41  \rangle} = \frac{g^2}{16}\frac{\langle  12  \rangle\langle  34  \rangle}{\langle  23  \rangle\langle  41  \rangle}\;.
\ee
In obtaining the above  we have used the anti-commutativity property
of  Grassmann variables and the embedding relation $\psi_i^I =
\theta^{\alpha I}\lambda_{i \alpha}$ to arrive at
\be
\int d^2\theta_1 \psi^1_i\psi^1_j = \int d^2\theta_1 \theta^{\alpha 1}\theta^{\beta 1}\lambda_{\alpha,i}\lambda_{\beta,j} = \frac{1}{2}\epsilon^{\alpha\beta}\lambda_{\alpha,i}\lambda_{\beta,j} = \frac{1}{2}\langle  j i \rangle \;.
\ee
On the spacetime side we have  contributions from two diagrams
\be
\begin{picture}(100,60)(-10,20)
\put(0,0){
\Text(-100,45)[l]{$\langle \phi, \phi, \phi^\dagger, \phi^\dagger\rangle_{4d} =$}
\SetColor{BrickRed}
\DashLine(10,80)(30,60){2}
\DashLine(50,80)(30,60){2}
\DashLine(10,10)(30,30){2}
\DashLine(50,10)(30,30){2}
\SetColor{Green}
\Photon(30,60)(30,30){3}{4}
\SetColor{Blue}
\Vertex(30,60){2}\Vertex(30,30){2}
\Text(5,85)[r]{$\phi_1$}
\Text(55,85)[l]{$\dagphi_4$}
\Text(25,55)[r]{$A_{\mu}$}
\Text(25,35)[r]{$A_{\nu}$}
\Text(5,5)[r]{$\phi_2$}
\Text(55,5)[l]{$\dagphi_3$}
\Text(95,45)[r]{$+$}
}
\end{picture}
\begin{picture}(80,80)(-10,10)
\put(15,0){
\SetColor{BrickRed}
\DashLine(10,60)(60,10){2}
\DashLine(10,10)(60,60){2}
\SetColor{Blue}
\Vertex(35,35){2}
\Text(5,65)[r]{$\phi_1$}
\Text(65,65)[l]{$\dagphi_4$}
\Text(5,5)[r]{$\phi_2$}
\Text(65,5)[l]{$\dagphi_3$}
}
\end{picture}
\ee
\vspace{.5cm}

\noindent and explicit calculation shows that the final result is $\langle \phi, \phi, \phi^\dagger, \phi^\dagger\rangle_{4d} = 32 i \langle \phi, \phi, \phi^\dagger, \phi^\dagger\rangle_{\mathrm{Twistor}}$ as claimed.

For the alternative ordering  $\langle \phi, \phi^\dagger, \phi, \phi^\dagger\rangle$   we have
\bea
\nn v_1(\phi) = \psi^1_1\psi^2_1 &,& v_2(\phi^\dagger) = \psi^3_2\psi^4_2\\
v_3 (\phi)=\psi^1_3\psi^2_3 &,& v_4(\phi^\dagger) = \psi^3_4\psi^4_4\;.
\eea
On the twistor side the amplitude is
\be
\langle\phi ,\phi^\dag, \phi, \phi^\dag\rangle_{\mathrm{Twistor}}=
g^2\int d^8\theta \frac{\psi^1_ 1\psi^2_1\psi^3_2\psi^4_2\psi^1_3\psi^2_3\psi^3_4\psi^4_4}
{\langle  12  \rangle \langle  23  \rangle\langle  34  \rangle\langle  41  \rangle} =\frac{g^2}{16}\frac{\langle 13\rangle^2 \langle 24 \rangle^2}
{\langle 12 \rangle \langle 23\rangle\langle 34\rangle\langle 41\rangle}
\ee
The spacetime side receives contributions from three Feynman diagrams
\be
\langle \phi, \phi^\dagger, \phi, \phi^\dagger\rangle_{4d}=
\begin{picture}(100,60)(-10,40)
\put(0,0){
\SetColor{BrickRed}
\DashLine(10,80)(30,60){2}
\DashLine(50,80)(30,60){2}
\DashLine(10,10)(30,30){2}
\DashLine(50,10)(30,30){2}
\SetColor{Green}
\Photon(30,60)(30,30){3}{4}
\SetColor{Blue}
\Vertex(30,60){2}\Vertex(30,30){2}
\Text(5,85)[r]{$\phi_1$}
\Text(55,85)[l]{$\phi^\dag_4$}
\Text(25,55)[r]{$A_{\mu}$}
\Text(25,35)[r]{$A_{\nu}$}
\Text(5,5)[r]{$\phi^\dag_2$}
\Text(55,5)[l]{$\phi_3$}
}
\end{picture}
+
\begin{picture}(120,60)(-10,30)
\put(0,0){
\SetColor{BrickRed}
\DashLine(10,50)(30,30){2}
\DashLine(60,30)(80,50){2}
\DashLine(10,10)(30,30){2}
\DashLine(60,30)(80,10){2}
\SetColor{Green}
\Photon(30,30)(60,30){3}{4}
\SetColor{Blue}
\Vertex(30,30){2}\Vertex(60,30){2}
\Text(5,55)[r]{$\phi_1$}
\Text(85,55)[l]{$\phi^\dag_4$}
\Text(35,25)[t]{$A_{\mu}$}
\Text(55,25)[t]{$A_{\nu}$}
\Text(5,5)[r]{$\phi^\dag_2$}
\Text(85,5)[l]{$\phi_3$}
}
\end{picture}
+
\begin{picture}(80,80)(-10,30)
\put(0,0){
\SetColor{BrickRed}
\DashLine(10,60)(60,10){2}
\DashLine(10,10)(60,60){2}
\SetColor{Blue}
\Vertex(35,35){2}
\Text(5,65)[r]{$\phi_1$}
\Text(65,65)[l]{$\phi^\dag_4$}
\Text(5,5)[r]{$\phi^\dag_2$}
\Text(65,5)[l]{$\phi_3$}
}
\end{picture}
\ee
\vspace{.5cm}

\noindent By explicit evaluation we once again find that $\langle \phi, \phi^\dagger, \phi, \phi^\dagger\rangle_{4d} = 32 i \langle \phi, \phi^\dagger, \phi, \phi^\dagger\rangle_{\mathrm{Twistor}}$.

\subsection{The amplitude  $\langle \eta_{A' }, \lambda^a , \bar\lambda^b, \bar\eta_{B'} \rangle$}

Let us also examine some more detailed results concerning  analytic amplitudes with  nontrivial dependence on the $\SU(2)$ indices $a,A$, which also involve external fundamental particles. As an example we consider $\langle \eta_{A' }, \lambda^a , \bar\lambda^b, \bar\eta_{B'} \rangle$. The wavefunctions can once again be read off from (\ref{VZ}) and (\ref{Q}) to give
\bea
\nn v_1(\eta_{A}) = \psi^A_1 &,& v_2(\lambda^a) = \epsilon_{da}\psi^d_2\\
v_3 (\tilde\lambda^b)=\epsilon_{cb}\psi^3_3\psi^4_3\psi^c_3 &,& v_4(\tilde\eta_{B}) = \psi^B_4\psi^1_4\psi^2_4\;.
\eea
The evaluation of the resulting integral  is highly simplified by the
use of various identities, collected in Appendix \ref{notation}. The answer is
\be
\langle \eta_{A }, \lambda^a , \tilde\lambda^b, \tilde\eta_{B} \rangle_{\mathrm{Twistor}} = g^2\epsilon_{cb}\epsilon_{da}\int d^8\theta \frac{\psi^A_
  1\psi^d_2\psi^3_3\psi^4_3\psi^c_3\psi^B_4\psi^1_4\psi^2_4}{\langle
  12  \rangle \langle  23  \rangle\langle  34  \rangle\langle  41
  \rangle} = \frac{g^2}{16}\epsilon_{ab}\epsilon^{AB}\left(
\frac{\langle  34  \rangle}{\langle  12  \rangle} + \frac{\langle  34  \rangle^2}{\langle  23  \rangle\langle  14  \rangle}\right)\;.
\ee
 On the other
hand, the diagrams contributing to the gauge theory calculation are
the following
\be
\langle \eta_{A' }, \lambda^a , \bar\lambda^b, \bar\eta_{B'} \rangle_{4d}=
\begin{picture}(100,60)(-10,40)
\put(0,0){
\SetColor{BrickRed}
\ArrowLine(10,80)(30,60)
\ArrowLine(30,60)(50,80)
\ArrowLine(30,30)(10,10)
\ArrowLine(50,10)(30,30)
\SetColor{Green}
\Photon(30,60)(30,30){3}{4}
\SetColor{Blue}
\Vertex(30,60){2}\Vertex(30,30){2}
\Text(5,85)[r]{$\eta_{A',1}$}
\Text(55,85)[l]{$\bar\eta_{B',4}$}
\Text(25,55)[r]{$A_{\mu}$}
\Text(25,35)[r]{$A_{\nu}$}
\Text(5,5)[r]{$\lambda^a_2$}
\Text(55,5)[l]{$\bar\lambda^b_3$}
}
\end{picture}
+
\begin{picture}(120,60)(-10,30)
\put(0,0){
\SetColor{BrickRed}
\ArrowLine(10,50)(30,30)
\ArrowLine(60,30)(80,50)
\ArrowLine(30,30)(10,10)
\ArrowLine(80,10)(60,30)
\DashLine(30,30)(60,30){2}
\SetColor{Blue}
\Vertex(30,30){2}\Vertex(60,30){2}
\Text(5,55)[r]{$\eta_{A',1}$}
\Text(85,55)[l]{$\bar\eta_{B',4}$}
\Text(35,25)[t]{$q^c_{\phantom c C}$}
\Text(55,25)[t]{$q^d_{\phantom d D}$}
\Text(5,5)[r]{$\lambda^a_2$}
\Text(85,5)[l]{$\bar\lambda^b_3$}
}
\end{picture}\\
\ee
\vspace{.5cm}

\noindent and explicit calculation  using  Feynman rules  leads to
$ \langle \eta_{A' }, \lambda^a , \bar\lambda^b, \bar\eta_{B'}
\rangle_{4d} = 32 i \langle \eta_{A }, \lambda^a , \tilde\lambda^b,
\tilde\eta_{B} \rangle_{\mathrm{Twistor}} $.

 Here we come to a crucial
point: When matching the spectra for the full
$N_f=4$ theory, we had already noticed that agreement could only be obtained if we decomposed
the global flavour index in terms of its special maximal subgroups
$\SO(8)\supset\SU(2)_{A'}\times\Sp(2)$ and then somehow related the
$\SU(2)_{A'}$ part to the flavour group for the antisymmetric
hypermultiplets $\SU(2)_A$. The requirement of matching  amplitudes
with external fundamental particles reaffirms this suggestion, since in
order to get agreement the two symmetries need to be identified! This
implies that the twistor string does not reproduce a gauge theory with
flavour group $\SO(8)$ but a theory which has had the latter
explicitly broken down to $\SU(2)\times\Sp(2)$. Moreover, this
$\SU(2)$ should then be realised geometrically on the gauge theory side;
recall that in the IIB description the flavour group for the antisymmetric hypermultiplet
fields was related to part of the rotations of the D3 worldvolume in
the transverse 6d space. The geometric realisation on the twistor
side is explicit and obvious in terms of the $\SU(2)_A$ symmetry
rotating the fermionic coordinates $\psi^A$. This result is quite
intriguing and we will briefly return to it in the conclusions.

\subsection{Further analytic amplitudes}
By now, the general strategy implemented for calculating 4--point
amplitudes on both sides of the correspondence should  be clear  to
the reader. Therefore, we will simply display the twistor
answer for several other analytic amplitudes which we have verified to match
those arising from the gauge theory calculation, up to the same relative
normalisation  factor of $32 i $. These amplitudes are
\bea
  \langle \lambda^a , \phi^\dagger , \bar\lambda^b , \phi \rangle &=&
  \frac{g^2}{16}\epsilon_{ab}\frac{\langle 23\rangle}{\langle 12\rangle} \\
\langle z^a_{\phantom a A} ,z^b_{\phantom b B} , z^c_{\phantom c C} ,
z^d_{\phantom d D} \rangle  &=& \frac{g^2}{16}\Big( - \frac{\langle  12 \rangle
\langle 34\rangle}{\langle 23\rangle\langle 14 \rangle }\epsilon_{ad}\epsilon_{bc}\epsilon^{AD}\epsilon^{BC}- \frac{\langle  14 \rangle
\langle 23\rangle }{\langle 12\rangle\langle 34
  \rangle }\epsilon_{ab}\epsilon_{cd}\epsilon^{AB}\epsilon^{CD} \\
\nn \quad &&
+\epsilon_{ab}\epsilon_{cd}\epsilon^{AD}\epsilon^{BC}+\epsilon_{ad}\epsilon_{bc}\epsilon^{AB}\epsilon^{CD}\Big)\\
\langle \phi^\dagger , z^a_{\phantom a A} ,z^b_{\phantom b B} , \phi \rangle
&=&\frac{g^2}{16} \frac{\langle  13 \rangle
\langle 24\rangle}{\langle 23\rangle\langle 14 \rangle }\epsilon_{ab}\epsilon^{AB} \\
\langle  z^a_{\phantom a A} ,\zeta_C , \bar\zeta_D , z^b_{\phantom b
  B} \rangle &=& -\frac{g^2}{16}\epsilon_{ab}\left(
\epsilon^{AB}\epsilon^{CD} \frac{\langle 13\rangle \langle 34
  \rangle}{\langle 23 \rangle\langle  14\rangle}+
\epsilon^{AC}\epsilon^{BD} \frac{\langle 13\rangle }{\langle
  12\rangle}\right)\;.
\eea
We recall that the partial amplitudes involving fundamental external
particles can be obtained directly from the antisymmetric ones by
(pair-wise) substitution of states. For example one has that
$\langle q^a_{\phantom a A} ,q^b_{\phantom b B} , q^c_{\phantom c C} ,
q^d_{\phantom d D} \rangle = \langle z^a_{\phantom a A} ,z^b_{\phantom b B} , z^c_{\phantom c C} ,z^d_{\phantom d D} \rangle  = \langle q^a_{\phantom a A} ,q^b_{\phantom b B} ,z^c_{\phantom c C} ,z^d_{\phantom d D} \rangle$ and so on.

These results strongly indicate that our proposed twistor duality for the $N_f=4$ theory,
as well as the assumption that (\ref{amp}) is applicable for amplitude
calculations, are valid. However, the structure of 4--point analytic amplitudes
is relatively trivial. A more concrete affirmation is given by examining and finding
agreement for 5--point  amplitudes. This would allow us to confidently state
that we are indeed considering the correct twistor string theory
dual. We have indeed
 explicitly checked this for the following two examples
\bea
\langle \lambda^a , z^b_{\phantom b B} , z^c_{\phantom c C}
,\lambda^d , \phi^\dagger \rangle &=& \frac{g^2}{16}\epsilon^{BC}\left(
\frac{\langle 25\rangle \langle 35 \rangle}{\langle 23 \rangle\langle
  45\rangle \langle 15 \rangle}\epsilon_{ad}\epsilon_{bc}  -
\frac{\langle 25\rangle \langle 35 \rangle \langle 14 \rangle}{\langle 12 \rangle\langle
  34\rangle \langle 45 \rangle \langle 15
  \rangle}\epsilon_{ab}\epsilon_{cd} \right)\\
\langle \phi , q^a_{\phantom a A} , q^b_{\phantom b B} , \eta_C ,
\eta_D\rangle &=& - \frac{g^2}{16}\epsilon_{ab}\left( \frac{\langle 13
  \rangle}{\langle 34 \rangle\langle 15\rangle} \epsilon^{AD}
\epsilon^{BC} -  \frac{\langle 13 \rangle\langle 25 \rangle}{\langle
  23 \rangle \langle 45\rangle \langle  15\rangle
}\epsilon^{AB} \epsilon^{CD} \right)
\; .
\eea
Once again, the results from the gauge theory side turn out to match those on
the twistor side up to the normalisation factor of $32i$.

\section{The $N_f = 2N $ theory}
We now turn our attention to another class of $\Ncal = 2 $ UV-finite gauge theories,
namely the theories with
gauge group $\SU(N)$ and flavour group $\SU(N_f)$, where $N_f=2N$. As discussed in the
introduction, this is the alternative way of extending the $\SU(2)$, $N_f=4$ theory of
Seiberg and Witten \cite{SeibergWitten9408}
beyond rank one. Here, we will identify the twistor string  dual to this  $N_f = 2N$ theory.
Since we have done most of the work in order to describe the $N_f = 4$ case, we will omit
some of the details in this case.

\subsection{Physical string theory description}
We will begin by reviewing the 10-dimensional  string theory description which realises
this gauge theory, in the same vein as for our $N_f = 4$ treatment. Unlike the previous case,
this theory does not have a natural connection to F--theory, but can instead be engineered as
the low energy worldvolume theory on a stack
 of $N$ fractional D3--branes probing the background generated by
 $N_f$ fractional D7--branes in Minkowski space with four
 orbifolded directions $\mathbb R^{1,5}\times \mathbb R^4/ \mathbb
 Z_2$. The latter are taken to be $(x^4,\ldots,x^7)$, with $\Zset_2$
acting on them as $(x^4,\ldots,x^7)\ra(-x^4,\ldots,-x^7)$.  We take the D3s
to lie along  $(x^0 , \ldots , x^3)$, and the D7s to be in $(x^0 , \ldots, x^7)$.
The D3--D7 system preserves 8
supercharges and the orbifold action has been chosen such that it does not break the supersymmetry
any further \cite{Polchinski00,GranaPolchinski01}. Once more, the 3--7 (7--3) strings provide the matter hypermultiplets transforming
in the fundamental (conjugate-fundamental) representation of the gauge group and in the
probe limit their $\SU(N_f)$ Chan-Paton index takes values in a global symmetry group. Similarly,
in this limit the `heavy' 7--7 strings decouple and one obtains a 4d $\Ncal = 2$, $\SU(N)$
gauge theory with $N_f$ fundamental hypermultiplets.

We are interested in the case where the D3s and D7s are located at the
same point in the transverse $(x^8, x^9)$ directions  (so there are no
masses for the matter fields) and where all D3s are coincident (that
is, no vevs). This is very reminiscent of the way we constructed the
$N_f = 4$ theory.  There are, however, some crucial differences:
Firstly, there is no orientifold plane in this case and hence no gauge
symmetry enhancement at any point on the moduli space; the gauge
groups corresponding to the open string degrees of freedom remain
$\SU(N)$. Secondly, the number of flavours corresponding to the
conformal point is chosen via a very different mechanism: On the
supergravity side the solution exhibits  a naked singularity, a
usual feature in the gravity description of non-conformal
theories. For the case of non-compact orbifolds, however, the
appearance of an enhan\c con \cite{Johnsonetal01} prevents the theory
from being trusted all the way to the singular point  since new, light
degrees of freedom appear at the enhan\c con radius. At that point,
the  SQCD energy scale diverges.  The excision of the region between
the enhan\c con radius and the naked singularity corresponds to
discarding energy scales where nonperturbative effects become
relevant. This also prevents one from obtaining a supergravity dual to
the gauge theory \`a la Maldacena.\footnote{The impossibility of obtaining a
supergravity dual even for the conformal $N_f=2N$ theory can also be seen by noting
(e.g. \cite{Hollowoodetal99}) that (unlike the $N_f=4$ theory) the two coefficients $a$ and $c$
of the four--dimensional anomaly are not equal to leading order in $1/N$, violating
a requirement of \cite{HenningsonSkenderis98,Gubser98}.} This system
therefore only
describes  the perturbative regime of the gauge theory, which is however
precisely the one that we want to reproduce from a twistor string
perspective. For $N_f  =2N$ the enhan\c con radius vanishes, the
gauge coupling stops running, and the theory sits at the
conformal point in its moduli space  \cite{Bertolinietal0107}.
In the following we will focus on this conformal $N_f=2N$ case.

Let us now take a look at the open string massless spectrum of the
theory. This is very similar to the one we studied for $N_f = 4$ and is
summarised in Table \ref{tablesu}. The orbifold projection  discards
the 3--3 open string modes responsible for the antisymmetric
hypermultiplets in the $N_f = 4$ theory. This  can be intuitively seen
from the inability of the fractional D3s to  move away from the
orbifold--fixed plane and therefore the antisymmetric hypermultiplet modes,
which were accounting for those degrees for freedom, are now absent.

\begin{table}[ht]
\begin{center}
\begin{tabular}{|c|c|c|c|c|c|c|}\hline
Component & SO(1,3) & $\SU(2)_a$ &$\SU(2)_A$ & $\Urm(1)_R$ & $\SU(N)$ &
$ \SU(2N)\!\!\times\!\!\Urm(1)$ \\ \hline
$A,G$ & $(2,2)$ & 1 & 1 & $0$ & $ N^2-1$ &1 \\
$\phi$& $(1,1)$ & 1 & 1  & $+2$ & $ N^2-1$ & 1\\
$\phi^\dagger$& $(1,1)$& 1 & 1  & $-2$ & $ N^2-1$ & 1\\
$\lambda_{\alpha, a}$& $(2,1)$ & 2 & 1 & $+1$ &  $N^2-1$ & 1 \\
$\bar{\lambda}_{\dot{\alpha}, a}$& $(1,2)$ & 2 & 1 & $-1$ &  $N^2-1$ & 1  \\ \hline
$q_{a}^I,q^{\dagger}_{aI}$& $(1,1)$& 2 & 1  & $0$ & $ N,\overline N$ &
$\overline{2 N}_{-1},2N_{+1} $\\
$\eta_{\alpha}^I,\bar\eta^{'I}_{\alpha}$& $(2,1)$ & 1 & 1 & $-1$ &  $N$ & $\overline{2N}_{-1}$ \\
$\bar{\eta}_{\dot{\alpha} I} , \eta'_{\dot{\alpha } I}$& $(1,2)$ & 1 &
1 & $+1$ &  $\overline N$ & $2N_{+1}$ \\ \hline
\end{tabular}
\end{center}
\caption{The on-shell field content of the $N_f=2N$ theory in component form. Once again,
the Lorentz representations are given in terms
of  $\SO(1,3)\rightarrow \SO(4)\sim\SU(2)_L\times\SU(2)_R$.
The fundamental fields carry an $\SU(2N)$ index $I = 1,\ldots , 2N$.}\label{tablesu}
\end{table}

Also note that no field transforms nontrivially under the $\SU(2)_A$. The reason we
include this symmetry in Table \ref{tablesu} is to precisely highlight the similarities and differences with the massless spectrum of the $N_f=4$ theory. The absence of the antisymmetric hypermultiplet is a sign that the discussion related to the geometric realisation of an $\SU(2)$ subgroup of the full flavour symmetry in the spacetime picture will not make an appearance in this context.

\subsection{The spacetime action}

We will now repeat the same steps as for the analysis of the $N_f = 4$ theory.
Without further delay, let us write down
the corresponding $\mathcal N =2 $ Lagrangean in terms of $\Ncal = 1$ superfields
\be\label{N2-2N}
\begin{split} \Lcal  = & \frac{1}{8\pi}\mathrm{Im}\; \mathrm{Tr}\left[ \tau  \left( \int
   d^2\theta\;  W^\alpha  W_{\alpha} + 2
  \int d^2 \theta d^2\bar{\theta}\; e^{2V}\Phi^{\dagger}e^{-2
   V} \Phi \right)\right]+
  \int d^2 \theta d^2\bar{\theta}\; Q^{\dagger I} e^{-2 V}Q_I \\
& +   \int d^2 \theta d^2\bar{\theta}\; Q'^I e^{2 V}Q'^\dagger_I
 +   \sqrt{2} \int d^2\theta\left(Q'^I  \Phi Q_I + h.c. \right) \; ,
\end{split}
\ee
where the  $I$s are now fundamental  $\SU(N_f)$ indices and
therefore $I = 1, \ldots , 2N$. The evaluation of the kinetic
part of the action will follow directly from the previous case by
setting the antisymmetric fields to zero and keeping in mind the new
global flavour group.  After expanding the superfields and performing
the Grassmann integration the  result reads
\be\label{neq2-2N}
\begin{split}
\Lcal  = &\frac{1}{g^2} \mathrm{Tr} \left( -\frac{1}{4}  F^2 +
 (D^\mu \phi)^{\dagger} (D_\mu \phi) - i \bar\lambda\Dslash\lambda - i
\bar\chi \Dslash\chi   - i
 \sqrt{2}\; [\lambda,\chi] \phi^\dagger - i \sqrt{2}\; [\bar{\lambda},
 \bar{\chi}] \phi
 \right)\\
&  +
(D^\mu q)^{\dagger I}(D_\mu q)_I +(D^\mu q')^I(D_\mu
 q')_I^{\dagger}  - i \bar \eta^I \Dslash
\eta_I  - i
\eta'^I \Dslash \bar\eta'_I-
i \sqrt 2 \; q^{\dagger I} \lambda \eta_I\\
&  + i\sqrt{2}\;
   \bar\eta^I\bar{\lambda} q_I  - i\sqrt{2}\;
  q'^I \bar \lambda \eta' _I + i\sqrt{2}\;
   \eta'^I \lambda q'^\dagger_I   - \sqrt 2 \left( \eta'^I\chi q_I +
\eta'^I \phi \eta_{I}   +   q'^I \chi
\eta_{I} \right)\\ &
 - \sqrt 2 \left( q^{\dagger I}\bar \chi \bar \eta'_I +
\bar\eta^I \phi^\dagger \bar\eta'_{I}   + \bar\eta^{I}  \bar\chi q'^\dagger_I
 \right)- V_S \; .
\end{split}
\ee
Once again $V_S$
denotes the scalar potential obtained by integrating out the auxiliary
${\sf F}$- and ${\sf D}$-terms, whose contributing  terms are now given by
\bea
 ({\sf F}_{q})_I^{ i} &=&-\sqrt 2\;
(\phi^\dagger)_{\phantom i j}^{i} q'^{\dagger j}_I\\ ({\sf F}_{
q'})^I_i &=& -\sqrt 2 \; q_{j}^{\dagger I } (\phi^\dagger)_{\phantom j
i}^j\\
({\sf F}_{\phi})^j_{\phantom j i} &=& -  g^2\sqrt 2 \; q_i^{\dagger I}
q'^{\dagger j}_I\\
{\sf D}^a &=& - \mathrm{Tr} \left(T^a[ \phi^\dagger , \phi]
\right)+g^2\left( q^{ \dagger
  I} T^a q_I -  q'^I T^a  q'^\dagger_I \right)\; ,
 \eea
where the $(T^a)_{\phantom i j}^i$'s are the generators of the
fundamental representation of $\SU(N)$. In the calculation of these terms we have,
in principle, the introduction of  $1/N$ contributions from
the coupling of the fundamental fields to the `photon'
\be
(T^a)^i_{\phantom i j} (T_a)^k_{\phantom k l} = \delta^i_l\delta^k_j
-\frac{1}{N} \delta^i_j\delta^k_l\; .
\ee
These, however, will  decouple  along with the rest of the
colour information during  the stripping process. We then perform the
field redefinitions (\ref{read}) and (\ref{refun}), and once again
combine fields in $\SU(2)_a$ doublets. The adjoint fermions are
redefined as in (\ref{doublets}), while for the fundamental
scalars we now have
\bea
\nn   (q^a)^i_{\phantom i I} = \doublet{q^i_{\phantom i I}}{-q'^{\dagger
 i}_{\phantom{\dagger i} I}} &,&  (q^\dagger_a)^I_{\phantom I i} =
 \left( q^{\dagger I}_{\phantom{\dagger I} i}, -q'^I_{\phantom I
 i}\right)\\
   (q^{\dagger a})^I_{\phantom I i} = \doublet{-q'^ I_{\phantom I i}}{-q^{\dagger
 I}_{\phantom{\dagger I} i }} &,&
 (q_a)^i_{\phantom i I} =
 \left( q'^{\dagger i}_{\phantom{\dagger i} I}, q^i_{\phantom i
 I}\right)\;.
\eea
The full action, including the quartic terms, now becomes
\be
\begin{split}
\mathcal L =  & \mathrm{Tr} \left[ -\frac{1}{2}  GF + \frac{1}{4}g^2G^2 +
 D\phi^{\dagger} D\phi + i \bar \lambda^a \Dslash \lambda_a
 -    \lambda^a\lambda_a \phi^{\dagger}  + 2  g^2
\bar \lambda^a \bar \lambda_a\phi  \right]-
 \eta'^I\phi\eta_I-2 g^2  \bar \eta^I\phi^\dagger\bar\eta'_I \\
&  -
 D q^{\dagger a I}Dq_{a I} - i \bar \eta^I \Dslash
\eta_I  - i
\eta'^I \Dslash \bar \eta'_I   +
   q^{\dagger a I} \lambda_a \eta_I -
   \eta'^I \lambda^a q_{a I} +2  g^2
   \bar\eta^I \bar{\lambda}^a q_{a I} - 2 g^2 q^{\dagger a I}  \bar \lambda_a
 \bar \eta'_I \\
  & \;-  \frac{g^2}{2}\mathrm{Tr} [\phi^\dagger, \phi]^2
  +  g^2 q^{\dagger a I}  \{\phi^\dagger , \phi \} q_{ a I} -
  \frac{g^2}{2} \left[ (q^{\dagger a I} q_{a J})(q^{\dagger
 b J} q_{ b I})  + (q^{\dagger I }_{a } q_{ b J} )(q^{\dagger a J} q^b_{I}) \right]  \\
&  +
  \frac{g^2}{2N} \left[ (q^{\dagger a I} q_{b I})(q^{\dagger
  J}_a q^b_{  J})  + (q^{\dagger  a I } q^b_{  I} )(q^{\dagger J}_b
 q_{ a J}) \right]\; .
\end{split}
\ee
In light of the twistor picture that we will discuss
in a moment, it seems natural to once again choose a special maximal
embedding of $\SU(2)$ into $\SU(2N)$, namely $\SU(2N)\supset
\SU(N)\times\SU(2)_{A'}$ and therefore we will decompose $I\rightarrow
K\otimes A'$, with $K = 1,\ldots , N$. Finally, after the appropriate
chiral rescalings (analogous to (\ref{read}), (\ref{refun})) the selfdual truncation of the
above is simply
\be\label{selfdualsu}
\begin{split}
\mathcal L =  & \mathrm{Tr} \left[ -\frac{1}{2}  GF  +
   D\phi^{\dagger} D\phi + i \bar \lambda^a \Dslash \lambda_a
 -    \lambda^a\lambda_a \phi^{\dagger}   \right] -   Dq^{\dagger a A'K} Dq_{a A'K} \\
&  - i \bar \eta^{A'K} \Dslash
\eta_{A'K}    - i
\eta'^{A'K} \Dslash \bar \eta'_{A'K}  - \eta'^{A'K}\phi\eta_{A'K} +q^{\dagger a A'K} \lambda_a \eta_{A'K} -
   \eta'^{A'K} \lambda^a q_{a A'K}
 \; . \end{split}
\ee

\subsection{The twistor action}
Let us now see how we can reproduce the spectrum of this $N_f=2N$ theory on
the twistor side and obtain the appropriate twistor action.
We will not provide exhaustive details for this
construction, since the arguments follow our previous analysis of the $N_f=4$ theory very closely.
To proceed, we simply orbifold two of the fermionic directions of $\Supertwistor$, namely
\be\label{orbidef}
\psi^a\ra \psi^a \quad,\quad \psi^A\ra -\psi^A
\ee
and place $N$ $\Dc$ branes spanning the bosonic and holomorphic fermionic directions,
as well as $N$ (rather than $2N$, which might seem more natural at first) $\Df$ branes on the
orbifold plane $\psi^3=\psi^4=0$ (as before,
this is loose language for ``branes satisfying DD boundary conditions in the
$\psi^3$,$\psi^4$ directions''). The orbifold action on the Chan-Paton
indices will again be given by $\gamma_c =\mathbb I_{N\times N} $ and $\gamma_f
=-\mathbb I_{N\times N}$. The invariant piece of the $c-c$ superfield $\Acal$ is
\be
\label{sungf}
\tilde\Acal  =
(A+\psi^a\lambda_a+\psi^1\psi^2\phi+\psi^3\psi^4\phi^\dagger
+\epsilon_{cd} \psi^3\psi^4\psi^c\tilde{\lambda}^d+
\psi^1\psi^2\psi^3\psi^4 G)\;,
\ee
which, via the arguments of the previous sections, will be mapped to the spectrum of an
$\Ncal =2$ vector multiplet in the adjoint of the gauge group $\SU(N)$.
Leaving aside the $f-f$ sector (the only
difference from the $N_f=4$ case being that the Chan--Paton indices will be
in $\SU(N)$, parametrised by $K=1\ldots N$) we will focus on the $c-f$ and $f-c$
strings. Arguing similarly to Section \ref{fundsec}, we find that the states
surviving the orbifold projection are now the following (0,1)--forms
\be
\mathcal{Q}^i_{\;K} = \psi^AQ^i_{AK}\qquad ,\qquad
\mathcal{Q}^{\dagger K}_{\;\;\;i} = \psi^A Q^{\dagger K}_{Ai}\;.
\ee
The $\dagger$ here simply denotes that these superfields transform in conjugate representations
of the gauge group $\SU(N)$, namely the fundamental and conjugate fundamental respectively.
We can further decompose $Q_A$ and $Q^{\dagger}_A$ into their components (suppressing
gauge indices from now on)
\bea
\nn Q_{AK} &=&
\eta_{AK}+ \psi^aq_{aAK}+ \psi^1 \psi^2
\tilde\eta'_{AK}\;, \\
Q^{\dagger K}_{A} &=&
\eta'^K_A+ \psi^a q^{\dagger K}_{aA}+ \psi^1 \psi^2
\tilde\eta^K_{A} \; .
\eea
The details related to identifying the BRST cohomology pertaining to the fundamental
superfields $\Qcal$, presented for the $N_f=4$ theory in Section \ref{Twistorsec},
will go through intact for this case as well.
The above expressions therefore provide the correct field content to reproduce
the spacetime spectrum for the fundamental hypermultiplets. It should now be
clear that $N$ $\Df$ branes suffice to provide the $2N$ hypermultiplets, although
in a form where the $\SU(2N)$ global group is not manifest.

The final twistor description is given by the hCS action
\be \label{Nf2Nsuperspace}
S =\int_{\Dc}\mathbf{\Omega} \wedge
\left(\half\mathrm{Tr}[\tilde{\mathcal A}\cdot\bar\partial \tilde{\mathcal A}+ \frac{2}{3} \tilde{\mathcal
A}\wedge\tilde{\mathcal A}\wedge\tilde{\mathcal A}]+ \mathcal Q^{\dagger K}\cdot
\bar \partial
\mathcal Q_{ K}+\mathcal Q^{\dagger K}\wedge\tilde{\mathcal
A} \wedge \mathcal Q_{ K} \right)\; ,
\ee
where $\tilde{\mathcal A}$ is as shown in (\ref{sungf}).
In component form this can be expanded into
\be \label{Nf2Ncomponent}
\begin{split}
  S_{hCS}= & \int_{\mathbb{C}\mathrm{P}^3}\mathbf{\Omega'}\wedge \big(\mathrm{Tr}[G\wedge {F}+
\phi^{\dag}\wedge\bar{D}\phi
-\tilde{\lambda}^a\wedge\bar{D}\lambda_a
+\lambda^a\wedge\lambda_a\wedge\phi^{\dag}] \\
 &  + \tilde \eta^{KA}\wedge\bar
 D\eta_{AK}+  \eta'^{KA}\wedge\bar
 D\tilde\eta'_{AK}-q^{\dagger aKA}\wedge\bar D q_{aAK}\\
 &  +
 \eta'^{KA}\wedge \phi\wedge \eta_{AK}-  q^{\dagger a KA
}\wedge \lambda_a \wedge \eta'_{AK}+ \eta'^{KA}\wedge \lambda_a
  \wedge q_{aAK}\big)\; .
\end{split}
\ee
The similarity with (\ref{selfdualsu}) is
obvious, once one identifies $A$ with $A'$. As we have already mentioned for
the $N_f = 4$ theory, we expect a nonlinear form of the Penrose transform to map the
above action to the selfdual truncation of the spacetime Lagrangean, given by
(\ref{selfdualsu}). As expected, we cannot assign a geometric meaning
to the spacetime $\SU(2)_{A'}$ in this case, even though the twistor
string description of $\SU(2)_A$ is explicitly geometric.
Note, however, that in the component action (\ref{Nf2Ncomponent}) (but
not in (\ref{Nf2Nsuperspace})) we can trivially
undo the $\SU(2N)\supset\SU(N)\times\SU(2)$ decomposition to exhibit the full
global flavour group $\SU(2N)\times\Urm(1)$. On the other hand, to apply the twistor
amplitude prescription (which explicitly involves the $\psi^A$ coordinates)
one is obliged to work with this symmetry non--manifest, and restore it at the end
by combining the relevant sets of amplitudes.\footnote{Instead, we chose to compute
gauge theory amplitudes in decomposed form and compare with the twistor results.}

Before proceeding to compare amplitudes, we should emphasise the similarities
between this construction for $N_f=2N$ and that for
the $N_f=4$ theory which we explored in Section \ref{Twistorsec}: The two
theories differ only by the presence of the orientifold and the number
of $\Df$ branes that are introduced. In the case of
rank one (where the orientifold imposes no condition, since $\SU(2)\cong\Sp(1)$)
they reduce to the same theory---the Seiberg--Witten $\SU(2)$ SYM with
four massless flavours. This simple picture is in contradistinction with the IIB
embeddings of these two theories, where (for instance) even the corresponding
orbifold actions are taken in different spacetime directions, and it is difficult
to see how they become equivalent for rank one. Presumably the twistor string
is able to be so concise in its description of this pair of theories because (unlike
their IIB string duals) it is only required to know about perturbative gauge theory
physics.

\subsection{Comparison of amplitudes}\label{Amplitudes2}

Finally, we move on to compare partial amplitudes on  both sides of the
correspondence. In fact, the similarity in  field content between the
$N_f=4$ and $N_f = 2N$  theories means that the partial amplitude
calculations are almost identical, since the only novelty, apart from the absence of the
antisymmetric hypermultiplet, is the behaviour of the fundamental
scalars and fermions due to the $\SU(N)$ gauge group. For example, it is easy
to see that the partial amplitude
involving  adjoint external particles is exactly the same
as for the $N_f = 4$ theory. Moreover, it is straightforward to replace the
 appropriate fundamental fields
and vertices to  find  the same agreement between the twistor
and spacetime results, including the relative normalisation factor of $32 i $.

As such, we only display two amplitudes. These involve fundamental external particles and,
at 4 and 5-point respectively, are
\bea
\langle q^{\dagger a}_{\phantom{\dagger a} A} ,q^b_{\phantom b B} , q^{\dagger c}_{\phantom{\dagger c} C} ,
q^d_{\phantom d D} \rangle _{\mathrm{Twistor}} &=& \frac{g^2}{16}\left( - \frac{\langle  12 \rangle
\langle 34\rangle}{\langle 23\rangle\langle 14 \rangle }\epsilon_{ad}\epsilon_{bc}\epsilon^{AD}\epsilon^{BC}- \frac{\langle  14 \rangle
\langle 23\rangle }{\langle 12\rangle\langle 34
  \rangle }\epsilon_{ab}\epsilon_{cd}\epsilon^{AB}\epsilon^{CD} \right.\\
\nn  &&
+\epsilon_{ab}\epsilon_{cd}\epsilon^{AD}\epsilon^{BC}+\epsilon_{ad}\epsilon_{bc}\epsilon^{AB}\epsilon^{CD}\Big)
\\
\langle \phi , q^a_{\phantom a A} , q^{\dagger b}_{\phantom{\dagger b} B} , \eta_C ,
\eta'_D\rangle_{\mathrm{Twistor}} &=& - \frac{g^2}{16}\epsilon_{ab}\left( \frac{\langle 13
  \rangle}{\langle 34 \rangle\langle 15\rangle} \epsilon^{AD}
\epsilon^{BC} -  \frac{\langle 13 \rangle\langle 25 \rangle}{\langle
  23 \rangle \langle 45\rangle \langle  15\rangle
}\epsilon^{AB} \epsilon^{CD} \right)\;.
\eea
A straightforward gauge theory calculation exactly reproduces these results.

\section{Conclusions and outlook}

In this paper we have extended the correspondence between 4d UV-finite
supersymmetric gauge theories and B--model twistor string theory at tree
level, by identifying the twistor string duals for theories containing
fundamental matter. These theories were $\Ncal  = 2 $, $\Sp(N)$
SYM with $N_f = 4$ and $\Ncal=2 $, $\SU(N)$ SYM with $N_f = 2N$
fundamental hypermultiplets, both sitting  at the superconformal point of
their moduli space. We initially studied the physical string
realisation of these theories and examined the open string massless
spectrum, which allowed us to properly identify all the symmetries of the
system. We then used this information to construct their
proper spacetime Lagrangean description. On the twistor side, we performed a
superorientifold and superorbifold projection respectively, which
yielded the non-fundamental part of the spectrum. The fundamental
degrees of freedom were introduced via new objects in the topological
B--model on supertwistor space, which we baptised  flavour-branes ($\Df$). These
wrap all the bosonic but only half of the holomorphic fermionic
directions spanned by Witten's Euclidean `D5'--branes providing the colour
degrees of freedom ($\Dc$).

We then proceeded to compare amplitudes on both sides of the proposed
correspondence. We found precise agreement for a number of 4-- and
5--point amplitudes, involving  external particles transforming in the
adjoint, fundamental, and, in the $N_f = 4$ case, antisymmetric
representations of the gauge group. These results provide strong
evidence for the robustness of the twistor string duals, and
even though we only calculated analytic (`MHV') processes in this work, we believe
that the agreement should continue to hold for tree level amplitudes supported
on holomorphic curves of  higher degree.

In the process of performing the identification between the two sides,
the embedding of the flavour-branes into the hCS
theory of the colour--branes forced us  to provide a geometric realisation for an
$\SU(2)$ subgroup of the flavour group, and in the $N_f = 4$ case to
identify that with the flavour symmetry of the antisymmetric
hypermultiplet fields. At first glance, this decomposition of the flavour group
might seem slightly \emph{ad hoc}; we could have chosen any
other subgroup which contains $\SU(2)$. However, our choice is
consistent with reproducing the same gauge group on the $\Df$ branes as the one appearing on
the $\Dc$ branes in the B--model, namely $\Sp(N)$ and $\SU(N)$ for the
two cases. The fact that, in the $N_f=4$ case, this decomposition leads to both kinds of branes
coming with the same type of gauge group ({\it i.e.} both symplectic) is
not unreasonable, if one remembers that they wrap the same number
of bosonic directions in $\Twistorspace$.

For the $N_f=4$ theory, we found that the twistor string
side actually describes a gauge theory with global flavour symmetry
broken down to $\SO(8)\rightarrow \SU(2)_A\times \Sp(2)$.
This was due to two unrelated (from the gauge theory point of view) $\SU(2)$ groups
being identified with the same geometric $\SU(2)_A$ on the twistor string side, and as such
the twistor string does not seem to describe precisely
the theory that we set off to recover. This could be so for a number of
reasons:
One possibility is clearly that we have not found the most generic twistor
string description of the $N_f=4$ theory, and that, despite the apparent
rigidity of our construction, further investigation might reveal a
way to disentangle these two symmetries.
A second possibility is that this is indeed the correct symmetry group of the IIB
setup once the effects of interactions between the fundamental and antisymmetric
hypermultiplet sectors are taken into account (recall that the claim that
the D3--D7 brane configuration accurately describes the $N_f=4$ theory is based mainly on
inspection of the spectrum). Checking this would entail establishing whether open
string interactions involving the antisymmetric hypermultiplet in the physical string picture preserve the global $\SO(8)$
flavour group or not.
A final possibility is that the twistor string actually maps to an enriched
version of the original physical brane construction. For example, this could arise
by taking the instantons on the D7
worldvolume theory away from the zero thickness limit, which,
if localised in the relative transverse
directions between the D3s and D7s, could break the global symmetry
precisely in the required fashion.\footnote{We would like to thank K.S. Narain
for suggesting such a possibility.} In this case, the mechanism leading to
the geometric interpretation of the $\SU(2)_A$ symmetry would be analogous
to the usual embedding of the gauge group into the spin connection.
 However, one is then forced to explain
why the twistor string only manages to capture the dynamics of this rather special
configuration, as well as to reconcile such a solution (which would seem to
move the theory towards the Higgs branch) with the apparently unbroken conformal
invariance.  It would be intriguing to uncover the answer to this
question, which we will, however, not address at present.
 We should emphasise that in the $N_f=2N$ theory the full flavour symmetry
is accurately (though not manifestly, given the decomposition
$\SU(2N)\rightarrow\SU(2)_A\times \SU(N)$) captured by the twistor side
and a spacetime geometric interpretation of the $\SU(2)_A$ on the IIB side
is not forced, essentially due to the absence of the antisymmetric hypermultiplet.

As discussed in the introduction, the main reason for studying twistor string duals
of finite theories is to potentially understand what, if anything, makes them
special on the twistor side. It is clear that generic non--finite theories are
not expected to have a dual with a $\Twistorspace$ component at the quantum
level, while the duals of the theories we have considered in this work should
have a $\Twistorspace$ description also at loop level. Unfortunately, since
our understanding of twistor string theory is confined to tree level, at this
stage we have
not been able to identify what is the distinguishing feature of our finite theories
as far as twistor strings are concerned. For example, for the theory considered
in Sections \ref{Nf4sec},\ref{Twistorsec} and \ref{Amplitudes}, we could just as well
have added one flavour-brane (and its mirror) instead of two, and the construction
would have worked
out in a very similar fashion, reproducing the amplitudes of an $\Ncal=2$ theory with
two (rather than four) fundamental hypermultiplets, clearly not a finite theory.
 The challenge, therefore, is to find a condition
(similar to the RR charge cancellation requirement which enforces $N_f=4$ on the physical
string side) which constrains the number of flavour-branes we can add to the
B--model on $\Supertwistor$.\footnote{We note that (bosonic)
topological string orientifolds in the twistor string context have been considered in
\cite{NeitzkeVafa04}. However, in that context the ensuing restriction on the number
of colour branes (and thus the rank of the $\Ncal=4$ SYM gauge group) was deemed
an unpleasant feature, and most consideration was given to orientifolds of
lower--dimensional subspaces of $\Supertwistor$. Perhaps the arguments in
\cite{NeitzkeVafa04} could be revisited with our current goal of restricting the
number of \emph{flavour} branes in mind.} An immediate obstacle is that our $\Df$
branes, whose number we would like to fix, have an $\Sp(2)$ gauge group, while in
the physical string context, orientifold planes leading to symplectic (rather
than orthogonal) groups on the corresponding branes have positive RR charges,
and thus are not relevant in situations where the total brane charge has to cancel.
However, in our topological context, this could perhaps be circumvented by recalling
the arguments of Vafa \cite{Vafa01} that topological anti--branes can be derived from
branes by formally taking $N\ra -N$. This, combined with the observation
\cite{Mkrtchyan81} that -- as far as gauge invariant quantities are
concerned -- in gauge theory $\Sp(N)$ can be thought of as $\SO(-N)$, indicates
that our $\Df$'s might be best thought of as anti--branes, whose negative `charge'
could potentially cancel that of the orientifold plane.
Similar comments apply to the $N_f=2N$ theory as
well, although the details will be different since in this case requiring finiteness
fixes the relative number of colour and flavour-branes rather than the absolute
number of $\Df$'s. Finding a mechanism that produces the above restrictions
should give considerable insight on how to properly complete the twistor
string description of finite gauge theories at the quantum level.\footnote{
And, applied in the other direction, might play a role in
establishing the UV--finiteness of other gauge or even gravity theories
admitting a twistor string description (a class which might, perhaps via a suitable
extension of the self--dual results of \cite{MasonWolf07},
potentially include $\Ncal = 8$ supergravity).}

 We should note that, although (as discussed above) our tree--level construction
(and the ensuing amplitude calculations) applies to gauge theories with
different numbers of flavours than those required for finiteness, for $\Sp(N)$ gauge
theories there seems to be a restriction to even numbers of flavours, since
we required (for $N=2$) the decomposition $4N\ra (2,2N)$ of the fundamental of $\SO(4N)$ under
$\SO(4N)\ra \SU(2)\times \Sp(N)$. At tree level (where finiteness constraints
should not arise) we might expect the twistor string to also describe theories
with \eg $N_f=3$, leading to an $\SO(6)$ flavour group, which would not fit in the
above framework. Perhaps a different
geometric embedding of the flavour-branes can account for such flavour groups.

Passing to other open directions suggested by our work,
it is interesting to remark (extending the discussion in \cite{SeibergWitten9408}
to higher rank) that the (massless as well as massive) $N_f=4$ theory is expected
to enjoy an analogue of the
$\SL(2,\Zset)$ Montonen--Olive symmetry of $\Ncal=4$ SYM, which combines
with $\mathrm{Spin}(8)$ triality to form the full
duality group of the theory. The $\SL(2,\Zset)$ duality of the $\Ncal=4$
theory motivated the authors of \cite{NeitzkeVafa04} to propose a strong--weak
duality relating the B--model with the A--model on the same (super) Calabi--Yau.
(Further discussion on the origin of this type of topological string duality
can be found in \cite{Nekrasovetal04}.)
It is intriguing to ask whether the duality group of the $N_f=4$ theory fits
within this framework, and therefore whether there exists an A--model version of
the setup we have constructed.  Also, the fact that the F--theory perspective
we reviewed in Section \ref{ftheorysec} provides a natural explanation of the duality
properties of the $N_f=4$ theory hints that perhaps a topological F--theory
\cite{Anguelovaetal04} point of view might provide some additional insight in this case.
Furthermore, given that the standard B--model $\Ncal=4$ SYM setup on $\Supertwistor$
has been conjecturally related (via the above S--duality plus mirror symmetry
arguments \cite{AganagicVafa04}) to a B--model on the superquadric
$\Lcal^{5|6}\in\Cset\mathrm{P}^{3|3} \times\Cset\mathrm{P}^{3|3}$ \cite{Witten78,Isenbergetal78},
it is natural to ask whether flavour-branes could also be incorporated in the latter geometry,
which should capture the dynamics of full (rather than self--dual) Yang--Mills theory without
the need for D1--instantons.

From a gauge theory point of view, one of the main interesting features of the
theories with fundamental matter we have considered is their richer vacuum structure
as compared to $\Ncal=4$ SYM, in particular the presence of Higgs branches. In the IIB
embeddings we have reviewed, this moduli space acquires geometric meaning,
in terms of the directions along which the various branes can be separated. Perhaps
the similarities of our constructions to the physical string realisations can
provide clues on how to move off the superconformal point from the twistor string
perspective as well.

In conclusion, we have demonstrated that the topological B--model
description of twistor strings is rich enough to accommodate
finite four--dimensional theories with fundamental matter, and that
the precise descriptions of these theories bear a strong resemblance to,
but also intriguing differences from, the standard embeddings of these theories
within physical string theory.
Apart from suggesting that a thorough analysis of boundary conditions and
associated D-branes for topological strings on supermanifolds (which was beyond
the scope of this work) would be a worthwhile enterprise,
we believe that our results reinforce the expectation that, by
deciphering the (still mysterious) connection between twistor and physical
strings, the current obstacles in establishing the twistor string duality at
the quantum level can eventually be overcome.

\vspace{1cm}

\noindent\textbf{Acknowledgements:}
 We would like to thank Lilia Anguelova, David Berman, Vincent Bouchard,
Cedric Delaunay, Dario Du\`o, Bobby Ezhuthachan, Antonio Grassi, Duc Ninh Le, K. S. Narain,
David Skinner, Gabriele Travaglini and Jun--Bao Wu for helpful
comments and discussions. J.B.  would like to acknowledge a Queen Mary
studentship and a Marie Curie Early Stage Training grant. The research of C.P. is supported by the
 Government of India. He is grateful to the organisers of Mideast'07 and Strings 2007 for
financial assistance while this work was being completed.  K.Z. is supported by PPARC through
the Special Programme Grant PP/C50426X/1 ``Gauge Theory, String Theory and Twistor Space
Techniques'' and would like to thank TIFR for generous
 hospitality during the latter part of this work.

\newpage
\begin{appendix}

\section{Notation and conventions}\label{notation}

In this short appendix we set up the notation and conventions used throughout this paper.

\paragraph{Spacetime:}

We take the signature of spacetime to be $(+---)$ and the raising and lowering of
spacetime spinor indices to be performed by
\bea
\nn \psi^\alpha = \epsilon^{\alpha\beta} \psi_\beta &,&
\psi_\alpha = \epsilon_{\alpha\beta} \psi^\beta\\
\bar \psi^{\dot\alpha} = \epsilon^{\dot\alpha\dot\beta} \bar
 \psi_{\dot\beta} &,&
\bar \psi_{\dot\alpha} = \epsilon_{\dot\alpha\dot\beta} \bar
 \psi^{\dot\beta}\; .
\eea
We also have the following relations between the superspace variables
\bea
\nn \theta^2 =\theta^\alpha\theta_\alpha = -2\theta^1\theta^2 &,&
  \theta^\alpha\theta^\beta =
 -\frac{1}{2}\epsilon^{\alpha\beta} \theta^2 \\
\bar\theta^2 = \bar
 \theta_{\dot\alpha} \bar \theta^{\dot\alpha} = 2\bar\theta_{\dot
 1}\bar\theta_{\dot 2} &,&
 \bar\theta_{\dot\alpha}\bar\theta_{\dot\beta} = -\frac{1}{2}
 \epsilon_{\dot\alpha\dot\beta} \bar\theta^2 \; .
\eea
The appropriate definitions for the $\epsilon$-tensors are
\be
\epsilon^{\alpha\beta} = \epsilon^{\dot\alpha\dot\beta } =
 \twobytwo{0}{1}{-1}{0} \; ,
\ee
where the above satisfy $\epsilon^{\alpha\beta}\epsilon_{\beta\gamma}=
 \delta^\alpha_\gamma$ and
 $\epsilon_{\dot\alpha\dot\beta}\epsilon^{\dot\beta\dot \gamma}=
 \delta_{\dot \alpha}^{\dot \gamma}$. Superspace integration then obeys
 \be
 \int\!  d\theta\, \theta= 1\qquad ,\qquad \int\!  d^2\theta\, \theta^\alpha\theta^\beta =  - \frac{1}{2} \epsilon^{\alpha\beta}
 \ee
and so on.
During the evaluation of amplitudes in twistor space, one also encounters
more complicated Grassmann integrals.  The following identities
dramatically simplify these superspace integrations (recall here
that $\psi^I=-\theta^{I}_{\alpha}\lambda^{\alpha}$)
\bea
\int d^4\theta \; \psi^a_i \psi^1_j \psi^2_j \psi^b_k & = &
\frac{1}{4} \epsilon^{ab}\langle ij \rangle\langle jk \rangle\\
\int d^4\theta \; \psi^A_i \psi^3_j \psi^4_j \psi^B_k & = &
\frac{1}{4} \epsilon^{AB}\langle ij \rangle\langle jk \rangle\\
\int d^4\theta \; \psi^a_i \psi^b_j \psi^c_k \psi^d_l & = &
\frac{1}{4}\left( \epsilon^{ad}\epsilon^{bc}\langle ij \rangle\langle kl
\rangle  -  \epsilon^{ab}\epsilon^{cd}\langle il \rangle\langle
jk\rangle\right)\\
\int d^4\theta \; \psi^A_i \psi^B_j \psi^C_k \psi^D_l & = &
\frac{1}{4}\left( \epsilon^{AD}\epsilon^{BC}\langle ij \rangle\langle kl
\rangle  -  \epsilon^{AB}\epsilon^{CD}\langle il \rangle\langle
jk\rangle\right)\;.
\eea
These expressions also lead to a useful $\epsilon$--tensor identity
\bea
\nn\int d^4\theta \; \psi^a_i \psi^b_j \psi^c_k \psi^d_l & = & - \int d^4\theta \; \psi^a_i  \psi^c_k\psi^b_j \psi^d_l\\
\nn \Rightarrow \frac{1}{4}\left( \epsilon^{ad}\epsilon^{bc}\langle ij \rangle\langle kl
\rangle  -  \epsilon^{ab}\epsilon^{cd}\langle il \rangle\langle
jk\rangle\right)&=& - \frac{1}{4}\left( \epsilon^{ad}\epsilon^{cb}\langle ik \rangle\langle jl
\rangle  -  \epsilon^{ac}\epsilon^{bd}\langle il \rangle\langle
kj\rangle\right)\\
\Rightarrow  \epsilon^{ad}\epsilon^{bc}+
\epsilon^{ab}\epsilon^{cd} &=&  \epsilon^{ac}\epsilon^{bd}\; .
\eea
 We  additionally make use of the following relations
\bea
\nn (\bar \sigma^\mu)^{\dot\alpha \alpha} =
 \epsilon^{\alpha\beta}\epsilon^{\dot\alpha\dot\beta}\sigma_{\beta\dot\beta}^\mu
 &,& \theta\sigma^\mu\bar\theta\theta\sigma^\nu\bar\theta = \frac{1}{2}\theta^2\bar\theta^2
 \eta^{\mu\nu}\\
\chi\sigma^\mu\bar\psi = -\bar\psi\bar\sigma^\mu \chi &,&
 (\chi\sigma^\mu\bar\psi)^\dagger  = \psi\sigma^\mu\bar\chi \; .
\eea

\paragraph{Gauge and flavour groups:}
The defining relation for elements of the $\Sp(N)$ algebra is that
\be
\label{spndefining}
M=-\Omega M^T \Omega^{-1}
\ee
for a hermitian matrix $M$, where $\Omega_{ij}$ is the invariant tensor of $\Sp(N)$.
The fundamental and conjugate-fundamental indices are then raised and lowered using
this tensor, which is defined via
\begin{displaymath}
\label{omega}
\begin{array}{ccc}
\Omega_{ij}=\Omega^{ij}=-(\Omega^{-1})^{ij}=i\sigma_2\otimes 1_{N\times N}\ ,&
\textrm{where}&
\sigma_2 = \twobytwo{0}{-i}{i}{0}
\end{array}
\end{displaymath}
and the indices are contracted following the `NW-SE' rule.
A useful property of matrices $M^i_{\;\;j}$ satisfying (\ref{spndefining}) is that
they become symmetric once their upper index is lowered using $\Omega_{ij}$.
Contraction of the invariant tensor gives $\Omega^{ik}\Omega_{kj}=-\delta^i_j$,
so that raising and lowering a
contracted $\Sp(N)$ index in a given expression results in the
appearance of an extra minus sign. In particular, in traces of products of
$\Sp(N)$ generators, the raising and lowering of
indices can be used to relate different permutations to each other which is
of importance when relating colour-stripped sub-amplitudes
to the full amplitudes. For example it is straightforward to see that
\begin{displaymath}
\begin{array}{ccc}
\Tr(T^aT^bT^c)=-\Tr(T^aT^cT^b)&
\textrm{and}&
\Tr(T^aT^bT^cT^d)=\Tr(T^aT^dT^cT^b)\; ,
\end{array}
\end{displaymath}
where $a,b,c,d$ here are adjoint indices. Furthermore, pseudoreality of the $\Sp(N)$
vector representation means that
fundamental and conjugate fundamental fields can be related  simply by
raising and lowering indices. Our assignment of signs for this is that
\be
\label{funantifun}
\Qcal_i=-\Omega_{ij}\Qcal^j\; .
\ee
\noindent Finally, as noted in equation (\ref{fierz}), the contraction of two $\Sp(N)$ generators gives
\be
\label{fierz2}
(T^a)^i_{\phantom i j} (T_a)^k_{\phantom k l} = \frac{1}{2}(
 \delta^i_l\delta^k_j - \Omega^{i
  k}\Omega_{j l} )\; .
\ee
More details on $\Sp(N)$ can be found, for instance, in \cite{Georgi}.

Because of the $\Sp(1)\cong
 \SU(2)$ isomorphism, the $\Sp(N)$ conventions above for the contraction of the invariant
tensor are the ones that we use for all
 other $\SU(2)$ symmetries (apart from the 4d Lorentz $\SU(2)$s discussed in the previous
section). In particular we take
\begin{eqnarray}
\epsilon_{ab}=\epsilon^{ab}=\twobytwo{0}{-1}{1}{0}
\end{eqnarray}
for raising and lowering $\SU(2)_a$ indices, which leads to $\gre_{ab}\gre^{bc}=-\delta_a^c$. Similar remarks
apply for $\SU(2)_A$.
Note, therefore, that for the Grassmann integration in supertwistor space these conventions imply
\be
 \int  d\psi \psi= 1\qquad ,\qquad \int  d^2\psi \psi^a\psi^b =  \frac{1}{2} \epsilon^{ab}\;.
 \ee
Our conventions for the $\SU(N)$ gauge and flavour indices are the usual ones to be found
in \emph{e.g.} \cite{PeskinSchroeder}.

\section{Feynman rules and useful identities }\label{Feynman}

In this appendix we present the Feynman rules and some related
identities, which we use for the calculation of
amplitudes in Sections \ref{Amplitudes} and \ref{Amplitudes2}.

\paragraph{Spinor identities}\mbox{}

In 4d, on-shell null momenta decompose in terms of two commuting, two-component,
positive and negative helicity spinors
$p_{\alpha\dot \alpha} = \lambda_\alpha \tilde\lambda_{\dot \alpha}$.
These are referred to as holomorphic and antiholomorphic spinors respectively
and we define the following inner products
\be
\lambda^\alpha \mu_{\alpha} = \langle\lambda \mu\rangle
\quad\textrm{and }\quad -\lt_{\dot \alpha}\tmu^{\dot \alpha} = [\lt\tmu]\;.
\ee
These products are antisymmetric so that $\langle \lambda \mu \rangle =  - \langle \mu \lambda \rangle$,
$[\lt \tilde \mu] = -[\tilde \mu \lt]$ and $\langle \lambda \lambda \rangle = [\lt \lt] = 0$.

One can switch between spinor helicity and Lorentz notations using the
generalised Pauli matrices $(\sigma^\mu)_{\alpha\dot \alpha}\equiv (1,\vec\sigma)$
and $(\bar\sigma^\mu)^{\dot \alpha \alpha }\equiv\epsilon^{\alpha\beta}\epsilon^{\dot\alpha \dot\beta}(\sigma^\mu)_{\beta\dot \beta}$ through
\be
q_{\alpha\dot \alpha} = {\sigma^\mu}_{\alpha\dot \alpha}q_\mu \quad ,\quad q^\mu
=\frac{1}{2}(\bar\sigma^\mu)^{\dot \alpha \alpha}q_{\alpha\dot \alpha}\;.
\ee
Some useful $\sigma$--matrix identities include
 \be
 (\sigma^\mu )_{\alpha\dot \alpha}(\bar\sigma_\mu )^{\dot\beta\beta}=
2\delta^\beta_{\alpha} \delta^{\dot\beta}_{\dot \alpha} \quad  , \quad
(\sigma^\mu)_{\alpha\dot \alpha}(\bar\sigma^\nu)^{\dot \alpha\alpha} = 2
\eta^{\mu\nu} \;.
\ee

The momentum inner product can be easily shown to be given by the expression
\be
 p\cdot q  = \frac{1}{2}\langle \lambda \mu \rangle [\lt \tmu]\; ,
\ee
which differs by a sign from the usual QCD literature but is in-line with
the majority of the twistor string literature.
Momentum conservation for an $n$--point amplitude can be implemented in
the spinor helicity formalism as\footnote{Here we use the common abbreviation
of $\langle\lambda_i\,\lambda_j\rangle=\langle i\,j\rangle$.}
\be
\sum_{i=1}^n\langle ji \rangle [ i k] =0 \;.
\ee
The Schouten identity is also extremely useful
\be
\langle ij \rangle \langle kl \rangle + \langle i k \rangle \langle l j \rangle + \langle i l \rangle\langle j k  \rangle = 0\; .
\ee

\paragraph{Feynman rules} \mbox{}

Here we list the Feynman rules for the $N_f = 4$ theory---the ones for the $N_f=2N$
theory can be obtained straightforwardly from these. In Table
\ref{Wavefunctions} we give the wavefunctions for external particles. Table
\ref{propagators} shows some examples of propagators, while Table
\ref{interactions} includes a few sample vertices.
The remaining vertices can of course be easily derived from the action.
The vertices for the $N_f = 2N$ theory are almost identical to the ones listed here. The
main differences are that the antisymmetric fields are absent in that case, and that the
fundamental scalars are complex as opposed to real fields.
In these expressions (as well as for our  amplitude calculations), all external momenta
are taken to be outgoing.
\begin{table}[ht]
\begin{center}
\begin{tabular}{|c|c|c|} \hline
Field & Helicity & Wave-function \\ \hline
Scalar & 0 & 1 \\
Fermion $i$& $+$ & $\lt_{i\,\dot \alpha}  =  - [i|$\\
Fermion $i$ & $-$ & $\lambda_i^{\alpha} = \langle i|$\\
Anti-fermion $j$& $+$ & $\lt_{j}^{\dot \alpha} =  |j] $\\
Anti-fermion $j$& $-$ & $\lambda_{j\,\alpha}  = |j\rangle$\\
Vector $p=\lambda\tilde{\lambda}$ & $+$ &
$\epsilon^+_{\alpha\dot \alpha} =\sqrt 2 \;
\frac{\mu_{\alpha}\lt_{\dot \alpha}}{\mu^\alpha \lambda_{\alpha}} = -\sqrt 2 \;
\frac{|\mu\rangle [ \lt |}{\langle \mu  \lambda\rangle}$ \\
Vector $p=\lambda\tilde{\lambda}$ & $-$ &
$\epsilon^-_{\alpha\dot \alpha} =\sqrt 2 \;
\frac{\lambda_{\alpha}\tmu_{\dot \alpha}}{\tmu_{ \dot \alpha} \lt^{\dot \alpha}}
= - \sqrt 2 \;
\frac{| \lambda\rangle[\tmu|}{[ \tmu \lt ]}$\\ \hline
\end{tabular}
\caption{Wavefunctions corresponding to outgoing external fields of given
helicity. Note that to define the vector wavefunctions we employ an arbitrary reference
vector $q=\mu\tilde{\mu}$.} \label{Wavefunctions}
\end{center}
\end{table}

\begin{table}[ht]
\begin{center}
\begin{tabular}{|c|c|c|} \hline
Field & Schematic form & Value\\ \hline
Adjoint scalar
&
\SetScale{0.9}
\begin{picture}(130,20)(0,30)
\thicklines
\SetColor{BrickRed}
\DashLine(30,30)(100,30){2}
\SetColor{Black}
\LongArrow(50,20)(78,20)
\Text(65,12)[pp]{\small$p$}
\Text(20,27)[b]{\small$\phi$}
\Text(105,27)[b]{\small$\phi^\dagger$}
\end{picture}
&
$\frac{i}{p^2}$\\
$q,z$ scalars
&
\SetScale{0.9}
\begin{picture}(130,30)(0,30)
\thicklines
\SetColor{BrickRed}
\DashLine(30,30)(100,30){2}
\SetColor{Black}
\LongArrow(50,20)(78,20)
\Text(65,12)[pp]{\small$p$}
\Text(20,30)[b]{\small$(q^a_{\phantom a A} , z^a_{\phantom a A})$}
\Text(110,30)[b]{\small$(q^b_{\phantom b B} , z^b_{\phantom b B})$}
\end{picture}
&
$\epsilon^{ab}\epsilon_{AB}\frac{i}{p^2}$\\
&& \\ \hline
Adjoint fermion
&
\SetScale{0.9}
\begin{picture}(130,30)(0,30)
\thicklines
\SetColor{BrickRed}
\ArrowLine(30,30)(100,30)
\SetColor{Black}
\LongArrow(50,20)(78,20)
\Text(65,12)[pp]{\small$p$}
\Text(20,27)[b]{\small$\lambda^a$}
\Text(105,27)[b]{\small$\bar\lambda^b$}
\end{picture}
&
$\epsilon^{ab}\frac{i p_{\alpha \dot \alpha}}{p^2}$\\
Adjoint antifermion
&
\SetScale{0.9}
\begin{picture}(130,30)(0,30)
\thicklines
\SetColor{BrickRed}
\ArrowLine(100,30)(30,30)
\SetColor{Black}
\LongArrow(50,20)(78,20)
\Text(65,12)[pp]{\small$p$}
\Text(20,27)[b]{\small$\bar\lambda^a$}
\Text(105,27)[b]{\small$\lambda^b$}
\end{picture}
&
$-\epsilon^{ab}\frac{i p^{\alpha \dot \alpha}}{p^2}$\\
$\eta,\zeta$ fermions
&
\SetScale{0.9}
\begin{picture}(130,30)(0,30)
\thicklines
\SetColor{BrickRed}
\ArrowLine(30,30)(100,30)
\SetColor{Black}
\LongArrow(50,20)(78,20)
\Text(65,12)[pp]{\small$p$}
\Text(20,30)[b]{\small$(\eta_A , \zeta_A)$}
\Text(110,27)[b]{\small$(\bar\eta_B , \bar\zeta_B)$}
\end{picture}
&
$\epsilon_{AB}\frac{i p_{\alpha \dot \alpha}}{p^2}$\\
&&\\ \hline
Vector
&
\SetScale{0.9}
\begin{picture}(130,30)(0,30)
\thicklines
\SetColor{Green}
\Photon(30,30)(100,30){4}{3}
\SetColor{Black}
\LongArrow(50,20)(78,20)
\Text(65,12)[pp]{\small$p$}
\Text(20,27)[b]{\small$A_\mu$}
\Text(105,27)[b]{\small$A_\nu$}
\end{picture}
&
$- ig^2\frac{\eta_{\mu\nu}}{p^2}\;.$\\
&&\\ \hline
\end{tabular}
\caption{Propagators for the various fields in our theory.} \label{propagators}
\end{center}
\end{table}

\newpage

\begin{table}[t]\label{vertices}
\begin{center}
\begin{tabular}{|c|c|} \hline
Schematic form & Value \\ \hline
\SetScale{0.4}
\begin{picture}(100,40)(0,10)
\put(0,0){
\SetColor{BrickRed}
\DashLine(130,80)(170,50){2}
\DashLine(170,50)(210,80){3}
\DashLine(130,20)(170,50){2}
\DashLine(170,50)(210,20){3}
\SetColor{Blue}
\Vertex(170,50){2}
\Text(30,30)[aa]{\tiny $(q^a_{\phantom a A} , z^a_{\phantom a A})$}
\Text(30,5)[aa]{\tiny$(q^b_{\phantom b B} , z^b_{\phantom b B})$}
\Text(90,30)[aa]{\tiny $\phi$}
\Text(90,5)[aa]{\tiny$\phi^\dagger$}
}
\end{picture}
& $i g^2 \epsilon_{ab}\epsilon^{AB}$\\
\SetScale{0.4}
\begin{picture}(160,40)(0,10)
\put(30,-5){
\SetColor{BrickRed}
\DashLine(130,80)(170,50){2}
\DashLine(170,50)(210,80){2}
\DashLine(130,20)(170,50){2}
\DashLine(170,50)(210,20){2}
\SetColor{Blue}
\Vertex(170,50){2}
\Text(30,30)[aa]{\tiny$(q^a_{\phantom a A} , z^a_{\phantom a A})$}
\Text(30,5)[aa]{\tiny$(q^b_{\phantom b B} , z^b_{\phantom b B})$}
\Text(110,30)[aa]{\tiny$(q^d_{\phantom d D} , z^d_{\phantom d D})$}
\Text(110,5)[aa]{\tiny$(q^c_{\phantom c C} , z^c_{\phantom c C})$}
}
\end{picture} &
\begin{tabular}{l}
$ i \left( 2 \epsilon_{ab}\epsilon_{cd}\epsilon^{AD}\epsilon^{BC}+\epsilon_{ad}\epsilon_{bc}\epsilon^{AD}\epsilon^{BC}\right. $\\
$\displaystyle \qquad\qquad \left.+2\epsilon_{ad} \epsilon_{bc}\epsilon^{AB}\epsilon^{CD}+\epsilon_{ab}\epsilon_{cd}\epsilon^{AB}\epsilon^{CD}\right) $\\
\end{tabular}\\
\SetScale{0.4}
\begin{picture}(100,40)(0,10)
\put(0,-5){
\SetColor{BrickRed}
\DashLine(130,80)(170,50){3}
\DashLine(170,50)(210,80){3}
\DashLine(130,20)(170,50){3}
\DashLine(170,50)(210,20){3}
\SetColor{Blue}
\Vertex(170,50){2}
\Text(45,35)[aa]{\tiny$\phi$}
\Text(45,6)[aa]{\tiny$\phi$}
\Text(90,35)[aa]{\tiny$\phi^\dagger$}
\Text(90,6)[aa]{\tiny$\phi^\dagger$}
}
\end{picture} &
$\qquad i g^2 $\\
\SetScale{0.4}
\begin{picture}(100,45)(0,10)
\put(-20,-10){
\SetColor{Green}
\Photon(215,100)(215,50){5}{3}
\SetColor{BrickRed}
\ArrowLine(215,50)(172,25)
\ArrowLine(258,25)(215,50)
\SetColor{Blue}
\Vertex(215,50){2}
\Text(82,45)[amu]{\tiny$A_\mu$}
\Text(64,14)[r]{\tiny$(\bar\eta_{B}, \bar\zeta_{B})$}
\Text(110,14)[l]{\tiny$(\eta_{A},\zeta_{A})$}
}
\end{picture} &
$\qquad -i \epsilon^{AB} \sigma^\mu$ \\
\SetScale{0.4}
\begin{picture}(100,45)(0,10)
\put(-20,-10){
\SetColor{BrickRed}
\DashLine(215,100)(215,50){2}
\ArrowLine(172,25)(215,50)
\ArrowLine(215,50)(258,25)
\SetColor{Blue}
\Vertex(215,50){2}
\Text(84,45)[amu]{\tiny$q^c_{\phantom c C }$}
\Text(64,14)[r]{\tiny$(\eta_{A}, \zeta_A)$}
\Text(110,14)[l]{\tiny$\lambda^a$}
}
\end{picture} &
$\qquad i \epsilon_{ac}\epsilon^{AC} $ \\
\SetScale{0.4}
\begin{picture}(100,40)(0,10)
\put(-20,-10){
\SetColor{BrickRed}
\DashLine(215,100)(215,50){2}
\ArrowLine(172,25)(215,50)
\ArrowLine(215,50)(258,25)
\SetColor{Blue}
\Vertex(215,50){2}
\Text(84,45)[amu]{\tiny$q^d_{\phantom d D}$}
\Text(64,14)[r]{\tiny$\bar\lambda^b$}
\Text(110,14)[l]{\tiny$(\bar\eta_{B},\bar\zeta_{B})$}
}
\end{picture} &
$\qquad   2 ig^2 \epsilon_{bd}\epsilon^{BD}$ \\
\SetScale{0.4}
\begin{picture}(100,50)(0,10)
\put(-20,-5){
\SetColor{Green}
\Photon(215,100)(215,50){5}{3}
\SetColor{BrickRed}
\DashLine(215,50)(172,25){2}
\DashLine(258,25)(215,50){2}
\SetColor{Blue}
\Vertex(215,50){2}
\Text(84,45)[amu]{\tiny$A_\mu$}
\Text(64,8)[r]{\tiny$\phi,1$}
\Text(110,8)[l]{\tiny$\phi^\dagger,2$}
}
\end{picture} &
$\qquad -i (p_2^\mu-p_1^\mu)$\\ &\\ \hline
\end{tabular}
\caption{Some of the interaction vertices of the $N_f=4$ theory.} \label{interactions}
\end{center}
\end{table}

\end{appendix}

\mbox{}
\newpage

\bibliography{TwistorRefs}
\bibliographystyle{JHEP1}

\end{document}